\pdfoutput=1
\documentclass[10pt]{IEEEtran}
\usepackage[dvipsnames, svgnames, x11names]{xcolor}
\usepackage{amsfonts}
\usepackage{amsmath}
\usepackage{graphicx}
\usepackage{float}
\usepackage{subcaption}
\usepackage{tikz}
\usepackage{ellipsis}
\usepackage{cite}
\usepackage{color}
\usepackage{algorithmic}
\usepackage{algorithm}
\usepackage[hidelinks]{hyperref}
\usepackage{caption}
\usepackage{placeins}
\usepackage{stfloats}
\usepackage{etoolbox}
\usepackage{microtype} 
\usepackage{booktabs}
\usepackage{multirow}
\usepackage{array}
\usepackage{amssymb}   
\usepackage{pifont}    
\usepackage{caption}

\newcommand{\yes}{\checkmark}
\newcommand{\no}{--}
\newtheorem{remark}{Remark}

\begin{document}
\title{Physical Layer Security for Movable Antenna-Aided Mixed-Field ISAC}

\author{Yun~Xiao,
        Enhao~Wang,~\IEEEmembership{Graduate Student Member,~IEEE,}
        Yunfei~Chen,~\IEEEmembership{Fellow,~IEEE,}
        Aissa~Ikhlef,~\IEEEmembership{Senior Member,~IEEE,}
        Zhuangkun Wei,
        and~Hongjian~Sun,~\IEEEmembership{Senior Member,~IEEE}
\thanks{Yun Xiao, Enhao Wang, Yunfei Chen, Aissa Ikhlef, Zhuangkun Wei, and Hongjian Sun are with the Department of Engineering, Durham University, Durham DH1 3LE, United Kingdom (e-mail: yun.xiao, enhao.wang, yunfei.chen, aissa.ikhlef, zhuangkun.wei, hongjian.sun@durham.ac.uk).}

}

\markboth{Physical Layer Security for Movable Antenna-Aided Mixed-Field ISAC}%
{Shell \MakeLowercase{\textit{et al.}}: Bare Demo of IEEEtran.cls for IEEE Journals}

\maketitle

\begin{abstract} 
Movable antenna (MA) can enhance the physical layer security (PLS) performance of integrated sensing and communications (ISAC). However, most existing works assume either near-field (NF) or far-field (FF) propagation models for both communication users (CUs) and targets. This paper considers a more practical scenario where multiple CUs, an eavesdropper (Eve), and multiple targets are in a mixed near- and far-field environment due to the changes in the propagation field caused by antenna position updates in MA with a unified array response and continuous NF/FF weights. To evaluate the PLS performance of MA-aided ISAC in mixed-field environments, we formulate an optimization problem to maximize the sum secrecy rate (SSR) of CUs by jointly designing the transmit beamforming and antenna positions at the transmitter.
The problem is highly non-convex, so a two-layer algorithm is proposed to iteratively solve the problem using the projected gradient ascent (PGA) method and the Lagrangian dual transform. Since the abrupt switch between the NF and the FF models during antenna movement can cause the objective function to oscillate during iteration, the unified array response is combined with the Barzilai-Borwein step, Armijo line search, and precise projection to ensure that the SSR does not decrease during the iteration and the objective function is differentiable on the Rayleigh boundary.
Numerical results show that the MA-ISAC using the mixed-field model achieves almost identical PLS performance to that of using the full NF model in terms of SSR, and both outperform the FF model. Using the mismatched FF model during antenna position updates results in a loss of MA gain, potentially making MA inferior to the FA-based system using the correct model.
\end{abstract}

\begin{IEEEkeywords}
Beamforming, integrated sensing and communications, mixed field, movable antennas, physical layer security.
\end{IEEEkeywords}

\IEEEpeerreviewmaketitle

\section{Introduction}
\IEEEPARstart{I}{ntegrated} sensing and communications (ISAC) has been regarded as a promising technology for the sixth-generation (6G) wireless communication networks by integrating the sensing and communication functions in a shared hardware platform to enable more efficient use of spectrum \cite{10418473}. With improved spectral and cost efficiency, ISAC is expected to support a wide range of applications, including the industrial Internet of Things (IIoT), environmental monitoring, smart homes, and vehicle-to-everything (V2X) \cite{11358925}. The massive growth in data transmission demands and the emergence of new ISAC applications have led to increased concern about information security \cite{11124234}. To prevent confidential information from being decoded by eavesdroppers (Eve), physical layer security (PLS) techniques have been widely employed in ISAC system designs \cite{11124234}.

On the other hand, 6G systems are expected to adopt extremely large antenna arrays and operate at extremely high frequencies (e.g., terahertz) to support high data rates. This significantly expands the near-field (NF) region determined by the Rayleigh distance \cite{11222687}. In the NF region, the electromagnetic (EM) field is more accurately characterized as a spherical wave, whereas in the conventional far-field (FF) region, it is typically approximated as a plane wave \cite{10944643}. Unlike the FF plane wave, which only contains angular information and steers the beam energy towards a specific direction, the NF spherical wave enables energy focusing at a specific location by exploiting additional distance information \cite{10944643}.

Due to the significantly expanded NF region in 6G, unlike most previous works on ISAC that often assume a pure NF model or a pure FF model for all devices, in practice, a mixed NF and FF scenario is more likely to arise, in which some targets or communication users (CUs) are located in the FF region, while others are located in the NF region. In this scenario, one may still use the NF model to describe both NF and FF, as FF is an approximation of NF when the distance is larger than the Rayleigh distance. The performance difference between applying the NF model and the FF model to FF CUs or targets may be negligible \cite{han2024cross}. However, the NF model requires significantly more parameters than the FF model. For example, a system with 1024 transmit and receive antennas has over $10^7$ parameters, resulting in highly complicated channel estimation for NF \cite{9508929}.
Therefore, it is still preferable to use a mixed-field model where NF users/targets are described by NF models and FF users/targets are described by FF models to save computational cost in ISAC.

Recently, flexible-antenna systems, such as fluid antennas, movable antennas (MAs), and pinching antennas, have also emerged as a key technology for enhancing ISAC performance by enabling dynamic adjustment of antenna positions at the transmitter (Tx) and/or receiver (Rx) \cite{9264694, 10286328, 11434944, 11169486}. Due to the high frequency and large array used, it is possible that the updates of antenna positions in MA may change the array aperture and thus the Rayleigh distance. This makes CUs and targets move between the FF and NF regions, and may cause the performance loss due to the mismatched propagation model, thereby reducing the potential benefits of the MA. Hence, it is important to consider propagation model changes during antenna position optimization, and this effect needs to be investigated in the PLS of MA-aided ISAC systems.

\subsection{Related Works}
PLS exploits the inherent randomness and spatial properties of wireless channels, combined with techniques such as beamforming (BF), interference mitigation, and artificial noise (AN), to provide reliable and secure transmission for ISAC networks \cite{11124234}. Existing studies on secure beamforming design for ISAC have primarily focused on pure NF or pure FF scenarios with fixed-position antenna (FA) arrays \cite{10540103, 10005137, 10227884, 11494878}. For FF ISAC systems, \cite{10540103} maximized the sensing signal-to-noise-ratio (SNR) of the target while guaranteeing the secrecy rate (SR) for all CUs in the presence of multiple Eves. Reference \cite{10005137} regarded the Eves as targets and showed that properly designed radar waveforms can serve as AN to suppress eavesdropping channels while providing additional degrees of freedom (DoF). With the aid of the AN, a sensing-assisted Eve estimation ISAC framework was developed in \cite{10227884} when the information of the Eve was unknown. The authors of \cite{11494878} maximized the sum secrecy rate (SSR) in secure multiple-input multiple-output (MIMO) integrated sensing and communications (ISAC) systems with imperfect channel state information (CSI) through joint beamforming and power allocation.

For NF ISAC systems, PLS can be further enhanced. For example, reference \cite{10436390} showed that, when the CU and the Eve were located at the same or similar spatial angles, applying the FF propagation model may cause performance loss due to the highly correlated channels between the CU and the Eve. The PLS of the NF ISAC based on non-orthogonal multiple access (NOMA) was investigated in \cite{10891691}, where the NF ISAC system achieved better security performance than its FF counterpart. In \cite{11096557}, the SSR and the Cram\'er-Rao bound (CRB) for target estimation were jointly optimized, and the NF propagation model enhanced both sensing accuracy and secure transmission performance.

The above studies assume that all CUs and targets follow either NF or FF propagation models. In practice, mixed-field occurs and ISAC in mixed-field has also been investigated \cite{10812003,xiao2024communication,11199344,10946468}. In \cite{10812003}, FF communications with an NF target and NF communications with an FF target were considered, while \cite{xiao2024communication} considered a communication-centric ISAC with multiple CUs in the mixed-field environment. These results showed the performance loss caused by the propagation model mismatch between NF and FF channels. Moreover, the results of \cite{11199344} revealed that, when the distance from the base station (BS) to the CU exceeds the Rayleigh distance, the performance degradation caused by employing the FF model becomes negligible relative to the NF model. Reference \cite{11185333} considered PLS for mixed-field communications, but without ISAC.

Recently, MA has emerged as a key technology to further improve ISAC performance \cite{10972180}. The MA-enabled FF ISAC was investigated in \cite{10839251,11436123,wu2026movableantennaenhancedisacoptimalantenna}. Specifically, \cite{10839251} combined bi-static sensing with a multi-user multiple-input single-output (MU-MISO) system, while \cite{11436123} maximized the minimum achievable CU rate under sensing CRB, transmit-power, and antenna-movement constraints. In addition, \cite{wu2026movableantennaenhancedisacoptimalantenna} jointly optimized MA trajectories and transmit beamforming using a globally optimal Branch-and-Bound (BnB) algorithm to minimize sensing beampattern mismatch subject to per-user signal-to-interference-plus-noise-ratio (SINR) constraints.

The MA-enabled NF ISAC was investigated in \cite{11075964}. The MA-aided mixed-field ISAC design without PLS was presented in \cite{11575300}, where the weighted sum of sensing and communications rate was maximized.
For secure communications without sensing, reference \cite{10416363} optimized the MA positions by projected gradient ascent (PGA) to maximize the secrecy rate in the FF region. To enhance the PLS in ISAC, reference \cite{11033708} considered MA and reconfigurable intelligent surface (RIS) for FF ISAC. The above works show that PLS can be further enhanced by using the MA. However, none of them has considered the PLS of MA-aided ISAC in the mixed-field scenario. Table \ref{tab:related_works} summarizes the difference between our work and the literature discussed.
\begin{table}[!t]
    \centering
    \caption{Comparison of Related Works}
    \label{tab:related_works}
    \renewcommand{\arraystretch}{1.05}
    \begin{tabular}{>{\centering\arraybackslash}m{2cm}
                    >{\centering\arraybackslash}m{1.4cm}
                    >{\centering\arraybackslash}m{1.0cm}
                    >{\centering\arraybackslash}m{1.0cm}
                    >{\centering\arraybackslash}m{1.0cm}}
        \toprule
        \textbf{Existing Works} & \textbf{Model} & \textbf{MA} & \textbf{PLS} & \textbf{ISAC} \\
        \midrule       
        \cite{10540103, 10005137, 10227884, 11494878}        & FF      & \no  & \yes & \yes \\
        \midrule
        \cite{10436390,10891691,11096557}        & NF      & \no  & \yes & \yes \\
        \midrule
        \cite{10812003,xiao2024communication,11199344,10946468}  & Mixed   & \no  & \no  & \yes \\
        \cite{11185333}        & Mixed   & \no  & \yes & \yes \\
        \cite{11575300}        & Mixed      & \yes & \no & \yes \\
        \midrule
        \cite{10839251,11436123,wu2026movableantennaenhancedisacoptimalantenna}        & FF      & \yes & \no  & \yes \\
        \cite{11075964}        & NF      & \yes & \no  & \yes \\
    \cite{10416363}       & FF      & \yes & \yes & \no  \\
        \cite{11033708}        & FF      & \yes & \yes & \yes \\
        \midrule
        \textbf{This work} & \textbf{Mixed} & \yes & \yes & \yes \\
        \bottomrule
    \end{tabular}
\end{table}
\subsection{Motivation and Contribution}
From the above discussions, several design challenges exist in the PLS of MA-aided ISAC systems in mixed-field scenarios. Firstly, it is unclear how the performance of MA-aided ISAC in the mixed-field is compared with those in conventional pure FF and pure NF scenarios, when the propagation model changes caused by antenna position updates is considered. Secondly, is it beneficial to use NF models for FF CUs/targets in mixed-field considering its increased computational complexity? 
Thirdly, how could model mismatch between NF and FF channels influence the PLS performance of ISAC, and how might this reduce the benefits of using MAs, as FA does not change antenna positions and therefore does not suffer from model mismatch caused by antenna position update? To answer these questions, this work considers a mixed-field ISAC system aided by MAs, as shown in Fig. \ref{fig:sys model}. Note that each of the core techniques used here, including mixed-field ISAC modeling, MA position optimization, PLS beamforming, PGA, and fractional programming (FP), has been studied individually in previous works but not jointly. In particular, the field transitions caused by antenna position updates have not been explored in prior studies, which either assume a fixed propagation regime or simply disregard field transitions. The main contributions are as follows:
\begin{itemize}
    \item We propose a mixed-field MA-aided ISAC system, where a dual-functional BS with MA-enabled transmitter simultaneously communicates with multiple single-antenna CUs and detects multiple sensing targets in the presence of a single-antenna Eve. The SSR is maximized under the constraints of the sensing signal-to-clutter-plus-noise ratio (SCNR) for all targets, the total transmit power, and antenna positions.
    \item A two-layer iterative algorithm is proposed to solve the optimization. First, the antenna positions are updated using the PGA method with Barzilai-Borwein (BB) step size, Armijo backtracking line search, and an explicit feasibility check on the sensing constraint, ensuring that the SSR is monotonically non-decreasing. Second, the beamformer is designed using the Lagrangian dual transform and convex approximation. Antenna positions and the beamformer are updated iteratively until the PLS performance converges to a desired threshold.
    \item During each antenna position update, the propagation models of the CUs, Eve, and targets are updated according to the new aperture and the new Rayleigh distance. To ensure a continuous transition between the NF and FF models, a unified array response with continuous, aperture-dependent NF weights is introduced, which includes pure NF and FF models as two special cases and ensures that the objective function is differentiable at the Rayleigh boundary, thereby avoiding the main source of oscillations during antenna position optimization.

    \item Numerical results show that the proposed algorithm using the mixed-field model achieves almost identical SSR to that using the full NF model at reduced complexity, while using the mismatched FF model for CUs/targets that move to the NF region after antenna movement results in a loss of MA gain, highlighting the importance of updating the propagation model during antenna position updates.
\end{itemize}

\subsection{Organizations and Notations}
The rest of this paper is structured as follows. The system model is presented in Section \ref{Sec:sys model}. Section \ref{Sec:BF design} formulates the optimization problem and proposes algorithms to achieve sub-optimal solutions for beamforming and antenna positions. Section \ref{Sec:results} provides numerical results and discussion, followed by the conclusion in Section \ref{Sec:conclusion}.

{\em Notations:} Complex vectors and matrices are represented by bold lowercase letters and bold uppercase letters, respectively. The ${\mathrm{tr(\cdot)}}$, ${( \cdot )^T}$, ${( \cdot )^H}$, and ${( \cdot )^*}$ denote the operators of trace, transpose, conjugate transpose, and conjugate, respectively. ${\mathbf{I}}$ denotes an identity matrix. ${\mathbb{C}^{m \times n}}$ denotes the set of $m \times n$ complex matrices. $\mathcal{R}$ denotes the real parts of parameters. $|s |$ denotes the magnitude of a complex value $s$. $\mathrm{E}\{  \cdot {\text{\} }}$ denotes the statistical expectation. $||\mathbf{s} ||$ denotes the Euclidean norm of a complex-valued vector $\mathbf{s}$.

\section{System Model}\label{Sec:sys model}
\begin{figure}
    \centering
    \includegraphics[width=0.35\textwidth]{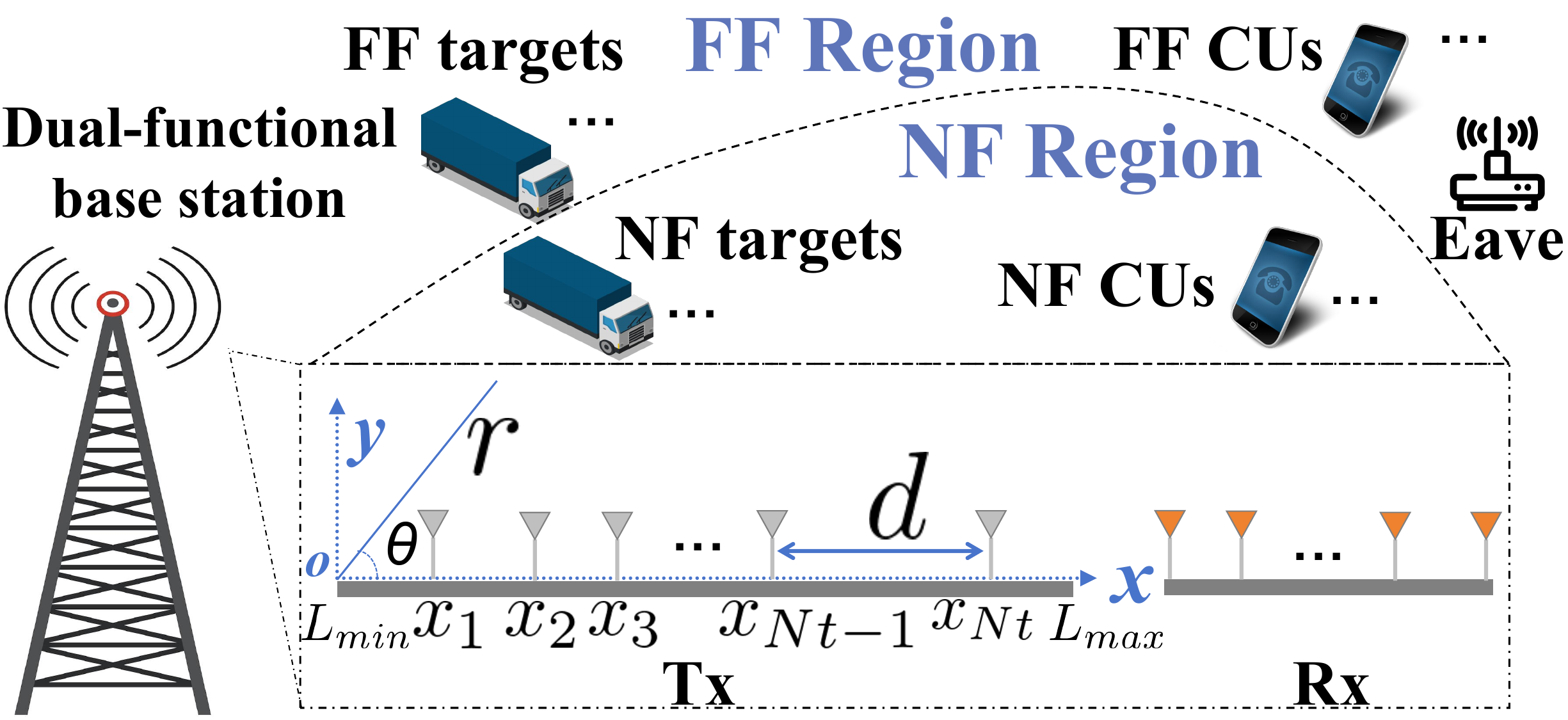}
    \captionsetup{font={small}} 
    \caption{System model.}
    \label{fig:sys model}
\end{figure}

An ISAC system with MA at the transmitter is shown in Fig. \ref{fig:sys model}, where the linear antenna array is adopted \cite{11222687}. The dual-functional BS is equipped with $N_{t}$ transmit antennas and $N_{r}$ receive antennas that simultaneously communicates with multiple CUs and detects multiple targets. MA is deployed only at the Tx for two reasons. First, the CUs and the Eve are considered as single-antenna devices, thus there are no receive-side antenna position optimization for them. Second, a receive-side MA would only benefit the sensing receiver, which is secondary to the PLS objective in this work. The placement of movable components at the transmitter is consistent with the transmitter-only MA-ISAC design in references \cite{10839251,11436123,wu2026movableantennaenhancedisacoptimalantenna,11075964}.
Assume that there are $M=M_1+M_2$ CUs in total, with $M_1$ single-antenna NF CUs denoted by $m_1\in \left \{ 1,...,M_1 \right \}$ and $M_2$ single-antenna FF CUs denoted by $m_2\in \left \{ 1,...,M_2 \right \}$. Similarly, there are $N=N_1+N_2$ targets in total, with $N_1$ NF targets denoted by $n_1\in \left \{ 1,...,N_1 \right \}$ and $N_2$ FF targets denoted by $n_2\in \left \{ 1,...,N_2 \right \}$. Note that the values of $M_1$, $M_2$, $N_1$, and $N_2$ may vary with the antenna array aperture for MA but remain fixed for FA. All CUs and targets are assumed to be static over $T$ time samples. The transmitted data from the BS is given by  
\begin{equation}\label{eq:transmitted signal}
    \textbf{d}\left [ t \right ] = \sum_{j=1}^{M}\textbf{b}_{j} d_{j}\left [ t \right ],
\end{equation}
where $t =1,\dots ,T$, $\textbf{b}_{j}\in  \mathbb{C}^{N_{t}\times 1}$ is the transmit beamforming vector of the $j$-th CU, and $d_{j}[t]\sim \mathcal{C} \mathcal{N}(0, 1) $ is the $j$-th CU's complex Gaussian-distributed data samples with mean $0$ and unit variance. The total transmit power can be expressed as
\begin{equation} \label{eq:Tx power}
\begin{aligned}
    \mathrm{E}(\left \| \textbf{d}[t] \right \|^{2}  )&=\sum_{j=1}^{M}\mathrm{tr}(\textbf{b}_{j}{\textbf{b}_{j}}^{H}).
\end{aligned}
\end{equation}

\subsection{Channel and Antenna Models}
Assume the uniform plane wave (UPW) model with linear phase and uniform power for FF and the non-uniform spherical wave (NUSW) model with non-linear phase and non-uniform power for NF \cite{10496996}. In this work, we adopt the  direction-dependent Rayleigh distance, which is defined as $d_{F}=\frac{2D^{2}\sin^2(\theta)}{\lambda}$ with the antenna array aperture $D$, CU/target angle $\theta$, and the signal wavelength $\lambda$ \cite{10496996}. Note that the change between FF and NF is not binary and there are other propagation models such as non-uniform plane wave (NUPW) and uniform spherical wave (USW)\cite{10496996}, but this Rayleigh distance is the most widely used boundary to distinguish the NF and FF regions. To avoid the abrupt switch between NF and FF regions, a unified array response that interpolates between the two limits is introduced at the end of this subsection. Let $r_j$ denote the distance of the $j$-th CU/target from the origin, and let $\theta_j \in [0,2\pi]$ represent its angle measured from the x-axis. The x-axis coordinates of the transmit antenna array are given by the set $\textbf{x}$, where each antenna can move between $L_{min}$ and $L_{max}$ with a minimum spacing of $d_{min}$ to avoid physical collision. 
The coordinate of the $n$-th array element is $[x_{n}, 0]$ for $n \in [1, \dots, N_{t}]$, and the array aperture is defined as $D = x_{N_t} - x_1$. The distance between the $n$-th array element and the $j$-th CU or target is then $r_j^{(n)} = \sqrt{r_j^{2} + x_{n}^{2} - 2 r_j x_{n} \cos(\theta_j)}$.
When $r_j^{(n)}<d_{F}$, the NF channel vector of the $j$-th CU/target is \cite{11283108}
\begin{equation}
    \left ( \textbf{h}_{\text{NF} }(\textbf{x}) \right )_j =\sqrt{\frac{\mathcal{K} }{1+\mathcal{K}} } \left ( \textbf{h}_{\text{LoS} }^{NF}(\textbf{x}) \right )_j  + \sqrt{\frac{1 }{1+\mathcal{K}} }\sum_{l=1}^{L} \left ( \textbf{h}_{\text{NLoS},l }^{NF} \right )_j ,
\end{equation}
where $\left (\textbf{h}_{\text{LoS} }^{NF}(\textbf{x})\right )_j\in \mathbb{C}^{N_{t}\times 1}$ is the NF line-of-sight (LoS) channel path, $\sum_{l=1}^{L} \left (\textbf{h}_{\text{NLoS},l }^{NF}\right )_j\in\mathbb{C}^{N_{t}\times 1}$ are the $L$ non-LoS
(NLoS) channel paths, $\mathcal{K}$ represents the Rician factor.
For the LoS path, we have
\begin{equation}\label{LoS NF channel}
    \left (\textbf{h}_{\text{LoS} }^{NF}(\textbf{x})\right )_j=\left (\boldsymbol{\beta}_{\text{LoS} }^{NF}\right )_j\odot \mathbf{a}\left (\textbf{x}, r_j,\theta_j \right ) ,
\end{equation}
where $\left (\boldsymbol{\beta}_{\text{LoS} }^{NF}\right )_j=\left [ \frac{\rho_{0}}{r_j^{(1)}}e^{-j\frac{2\pi}{\lambda}r_j},\dots, \frac{\rho_{0}}{r_j^{(N_t)}}e^{-j\frac{2\pi}{\lambda}r_j}\right ]^T$ consists of complex-valued channel gains for each antenna, $\odot$ is the Hadamard product, $\rho_{0}=\frac{\lambda}{4\pi}$ is the reference path gain with given wavelength $\lambda$ \cite{11199344}, $\mathbf{a}\left ( \textbf{x},r_j,\theta_j\right )$ is the NF beam focusing vector with
\begin{equation}\label{beam focusing}
    \mathbf{a}\left (\textbf{x}, r_j,\theta_j \right ) =\left [ e^{-j\frac{2\pi}{\lambda}\left ( r_j^{(1)}-r_j \right )  },\dots , e^{-j\frac{2\pi}{\lambda}\left ( r_j^{(N_t) }-r_j \right )  } \right ]^{T}.
\end{equation}
When $r_j^{(n)}>d_{F}$, the FF channel of the $j$-th CU/target is
\begin{equation}\textcolor{red}{\label{FF total channel}}
    \left (\textbf{h}_{\text{FF} }(\textbf{x})\right )_j=\sqrt{\frac{\mathcal{K} }{1+\mathcal{K}} } \left (\textbf{h}_{\text{LoS} }^{FF}(\textbf{x})\right )_j + \sqrt{\frac{1 }{1+\mathcal{K}} }\sum_{l=1}^{L} \left (\textbf{h}_{\text{NLoS},l }^{FF}\right )_j,
\end{equation}
where $\left (\textbf{h}_{\text{LoS} }^{FF}(\textbf{x})\right )_j\in \mathbb{C}^{N_{t}\times 1}$ and $\left (\textbf{h}_{\text{NLoS}}^{FF}\right )_j\in \mathbb{C}^{N_{t}\times 1}$ are the LoS and NLoS channels in FF, respectively. For the FF LoS channel, it can be derived by approximating the NF channel using the Taylor series expansion as \cite{10496996}
\begin{equation}\label{LoS FF channel}
    \left (\mathbf{h}_{\text{LoS}}^{FF}(\textbf{x})\right )_j=\left (\boldsymbol{\beta}_{\text{LoS}}^{FF}\right )_j\odot  \mathbf{a}(\textbf{x},\theta_j)
\end{equation}
where $\left (\boldsymbol{\beta}_{\text{LoS}}^{FF}\right )_j=\left [ \frac{\rho_{0}}{r_j}e^{-j\frac{2\pi}{\lambda}r_j},\dots,\frac{\rho_{0}}{r_j}e^{-j\frac{2\pi}{\lambda}r_j} \right ]^{T} $ is the $N_{t}\times 1$ vector of channel gains for all antennas, $\textbf{a}(\textbf{x},\theta_j)$ is the FF beam steering vector with
\begin{equation}\label{beam steering}
    \textbf{a}(\textbf{x},\theta_j)= [ e^{j\frac{2\pi}{\lambda}x_{1}\cos(\theta_j )},
\dots,e^{j\frac{2\pi}{\lambda}x_{N_t}\cos(\theta_j )} ] ^{T}.
\end{equation}
Note that the FF model in (\ref{LoS FF channel}) is an approximation of the NF model in (\ref{LoS NF channel}). Hence, in the mixed-field scenario, one may use the NF model in (\ref{LoS NF channel}) to describe both NF and FF channels. However, NF incurs high computational cost. As can be seen from (\ref{LoS NF channel}), NF channel has a lot of channel parameters to estimate, especially when $N_t$ is very large. Thus, it is still preferable to use NF model to describe NF channels and FF model to describe FF channels in a mixed-field scenario to save computational cost.
\begin{remark}\label{rem:nlos}Each NLoS path has the same form as the LoS path, but evaluated at a slightly different angle $\theta_j+\Delta\theta_l$ and distance $r_j+\Delta r_l$, representing scattering near the $j$-th user/target. $\Delta\theta_l\sim\mathcal{N}(0,\sigma_\theta^2)$ and $\Delta r_l\sim\mathcal{U}(0,\Delta r_{\max})$ are random and change with each channel realization. Since these offsets are random scattering effects rather than fixed functions of position $\mathbf{x}$, the position gradient in Section \ref{Sec:BF design}-C uses only the LoS path.\end{remark}

\textit{Unified NF/FF array response:} The unified NF/FF model has been used for codebook design \cite{chen2024unifying} and hybrid field source localization \cite{fang2024threedimensional}, here we adopt a model to ensure that the position optimization objective is differentiable on the NF/FF boundary. The NF and FF models differ only by a phase correction term $\delta_{n,j}$ that captures wavefront curvature, which is given by
\begin{equation}\label{eq:residual}
    \delta_{n,j}=\frac{2\pi}{\lambda}\left(r_j^{(n)}-r_j+x_n\cos(\theta_j)\right)\approx\frac{2\pi}{\lambda}\frac{x_n^2\sin^2(\theta_j)}{2r_j}.
\end{equation}
This term grows continuously with $D^2/(\lambda r_j)$, so the propagation model transition at the NF/FF boundary is smooth with no abrupt switch. We introduce a continuous weight $\omega_j\in[0,1]$ for each device and define the unified array response as
\begin{equation}\label{eq:unified response}
    \left[\mathbf{a}\left(\mathbf{x},r_j,\theta_j,\omega_j\right)\right]_n=e^{j\frac{2\pi}{\lambda}x_n\cos(\theta_j)}\,e^{-j\omega_j\delta_{n,j}},\ n=1,\dots,N_t,
\end{equation}
where $\omega_j=1$ gives the pure NF model and $\omega_j=0$ gives the pure FF model. Channel gains are similarly formulated as
$\left(\boldsymbol{\beta}_{\text{LoS}}(\omega_j)\right)_j=\omega_j\left(\boldsymbol{\beta}_{\text{LoS}}^{NF}\right)_j+(1-\omega_j)\left(\boldsymbol{\beta}_{\text{LoS}}^{FF}\right)_j$.
The weight is determined by the ratio $\xi_j=2D^2\sin^2(\theta_j)/(\lambda r_j)$, which measures how deep a device is in the NF region. We use a smooth ramp function to determine $\omega_j$, which is given by
\begin{equation}\label{eq:omega}
    \omega_j=S\!\left(\frac{\xi_j-\xi_{lo}}{\xi_{hi}-\xi_{lo}}\right),\quad S(t)=\begin{cases}0, & t\le 0,\\ 3t^2-2t^3, & 0<t<1,\\ 1, & t\ge 1,\end{cases}
\end{equation}
where $[\xi_{lo},\xi_{hi}]$ defines the transition window around the Rayleigh boundary $\xi_j=1$. Since $\omega_j$ depends on $\mathbf{x}$ only through the aperture $D=x_{N_t}-x_1$, the unified channel remains differentiable, enabling gradient-based optimization in Section \ref{Sec:BF design}-C.

\subsection{Communications Performance Metric}
Given the transmitted signal $\textbf{d}\left [ t \right ]$, the received signal at the $j$-th CU is given by
\begin{equation}
    y_{j}\left [ t \right ]=\mathbf{h}_{j}^{T}(\textbf{x})\mathbf{b}_{j} d_{j}\left [ t \right ]+\mathbf{h}_{j}^{T}(\textbf{x})\sum_{k\ne j}^{}\mathbf{b}_{k} d_{k}\left [ t \right ]+z_{j}\left [ t \right ],
\end{equation}
where $\mathbf{h}_{j}(\textbf{x})\in\left \{ (\mathbf{h}_{\text{NF}}(\textbf{x}))_{j},(\mathbf{h}_{\text{FF}}(\textbf{x}))_{j} \right \} $ denotes the $N_{t}\times 1$ channel vector of the $j$-th CU, and $z_{j}[t]\sim \mathcal{CN}\left ( 0,\sigma_{j}^2 \right )$ represents the additive white Gaussian noise (AWGN) at the $j$-th CU. Similarly, the received signal at the Eve is given by
\begin{equation}
    y_{e}\left [ t \right ]=\mathbf{h}_{e}^{T}(\textbf{x})\sum_{k=1}^{M}\mathbf{b}_{k} d_{k}\left [ t \right ]+z_{e}\left [ t \right ],
\end{equation}
where $\mathbf{h}_{e}(\textbf{x})\in\mathbb{C}^{N_{t}\times 1}$ is the channel vector of the Eve, and $z_{e}\left [ t \right ]\sim \mathcal{CN}\left ( 0,\sigma_{e}^2 \right )$ is the AWGN at the Eve.
Let $\boldsymbol{\mathcal{H}}_{j}(\textbf{x})=\mathbf{h}_{j}(\textbf{x})\mathbf{h}_{j}^{H}(\textbf{x})$. The SINR at the $j$-th CU is then given by
\begin{equation}\label{SINR}   \text{SINR}_{j}=\frac{\mathbf{b}_{j}^{H} \boldsymbol{\mathcal{H}}_{j}(\mathbf{x}) \mathbf{b}_{j}}
{\sum_{k=1,\,k\ne j}^{M} \mathbf{b}_{k}^{H}\boldsymbol{\mathcal{H}}_{j}(\mathbf{x})\mathbf{b}_{k} + \sigma _{j}^{2}},\forall j.
\end{equation}
Let $\boldsymbol{\mathcal{H}}_{e}(\textbf{x})=\mathbf{h}_{e}(\textbf{x})\mathbf{h}_{e}^{H}(\textbf{x})$. The SINR at the Eve for the $j$-th CU is given by
\begin{equation}\label{SINR_e}
    \text{SINR}_{e,j}=\frac{\mathbf{b}_{j}^{H} \boldsymbol{\mathcal{H}}_{e}(\mathbf{x})\mathbf{b}_{j}}
{\sum_{k=1,\,k\ne j}^{M} \mathbf{b}_{k}^{H} \boldsymbol{\mathcal{H}}_{e}(\mathbf{x})\mathbf{b}_{k} + \sigma _{e}^{2}}, \forall j.
\end{equation}
Then, the SSR can be derived as
\begin{equation}
\begin{aligned}
SSR =& \sum_{j=1}^{M} \Bigg[
\log\!\left(1+ \frac{\mathbf{b}_{j}^{H} \boldsymbol{\mathcal{H}}_{j}(\mathbf{x}) \mathbf{b}_{j}}
{\sum_{k=1,\,k\ne j}^{M} \mathbf{b}_{k}^{H}\boldsymbol{\mathcal{H}}_{j}(\mathbf{x})\mathbf{b}_{k} + \sigma _{j}^{2}}\right) \\
&- \log\!\left( 1+\frac{\mathbf{b}_{j}^{H} \boldsymbol{\mathcal{H}}_{e}(\mathbf{x})\mathbf{b}_{j}}
{\sum_{k=1,\,k\ne j}^{M} \mathbf{b}_{k}^{H} \boldsymbol{\mathcal{H}}_{e}(\mathbf{x})\mathbf{b}_{k} + \sigma _{e}^{2}} \right)
\Bigg]^{+},
\end{aligned}
\end{equation}
where $[x]^{+}=\max\{x,0\}$.
\subsection{Sensing Performance Metric}
Given $\textbf{d}\left [ t \right ]$ in (\ref{eq:transmitted signal}), the signal at the sensing receiver is
\begin{equation} \label{sensing received signal}
\begin{aligned}
  y_{s,i}[t]  
       &=\underbrace{ \textbf{f}_{i}^{H}\sum_{i=1}^{N}(\boldsymbol{\beta}_{i}\odot\textbf{A}_{i})\textbf{d}[t]}_{\text{sensing signals}}+ \underbrace{\textbf{f}_{i}^{H}\sum_{c=1}^{N_c}(\boldsymbol{\beta}_{c}\odot\textbf{A}_{c})\textbf{d}[t]}_{\text{clutters}} +\underbrace{\textbf{f}_{i}^{H}\textbf{z}_{s}[t]}_{\text{AWGN}}.
\end{aligned} 
\end{equation} 
where $N_c$ denotes the number of clutters, the matrices $\textbf{A}_{i}=\textbf{a}_{ri}\textbf{a}_{ti}^{T}$ and $\textbf{A}_{c}=\textbf{a}_{rc}\textbf{a}_{tc}^{T}$ represent the beam steering or focusing matrices, with $\left \{ \textbf{a}_{ri},\textbf{a}_{rc} \right \}$ and $\left \{ \textbf{a}_{ti},\textbf{a}_{tc} \right \}$ being the receive and transmit beam steering vectors in (\ref{beam steering}) or focusing vectors in (\ref{beam focusing}), respectively. The AWGN is given by $\textbf{z}_{s}[t]\sim \mathcal{CN}\left ( 0,\sigma_{s}^2\textbf{I} \right )$, and the complex channel gains of the $i$-th target and $c$-th clutter are denoted by $\boldsymbol{\beta}_{i}$ and $\boldsymbol{\beta}_{c}$, respectively. Additionally, $\textbf{f}_{i}\in\mathbb{C}^{N_{r}\times 1 }$ is the receive beamforming vector for the $i$-th target. By solving the minimum variance distortionless response (MVDR) problem, its optimal value is obtained as \cite{11199344}
\begin{equation}\label{eq:mvdr}
    \textbf{f}_{i}=\frac{{\textstyle \sum_{j=1}^{M}}\left ( \textbf{R}_{i}+\textbf{R}_{c}+\sigma_{s}^{2}\textbf{I} \right )^{-1}\textbf{A}_{i}\textbf{b}_{j} }
{{\textstyle \sum_{j=1}^{M}}\textbf{b}_{j}^{H}\textbf{A}_{i}^{H}\left ( \textbf{R}_{i}+\textbf{R}_{c}+\sigma_{s}^{2}\textbf{I}\right )^{-1}\textbf{A}_{i}\textbf{b}_{j}},
\end{equation} 
where $\textbf{R}_{i}={\textstyle \sum_{u\ne i}^{N}} (\boldsymbol{\beta}_{u}\odot \textbf{A}_{u})\left ( \sum_{j=1}^{M}\textbf{b}_{j}\textbf{b}_{j}^{H} \right ) (\boldsymbol{\beta}_{u}\odot \textbf{A}_{u})^{H}$, $\textbf{R}_{c}=\sum_{c=1}^{N_c} (\beta _{c} \odot \textbf{A}_{c}) \left ( \sum_{j=1}^{M}\textbf{b}_{j}\textbf{b}_{j}^{H} \right ) (\beta _{c} \odot \textbf{A}_{c})^{H}$, and $\sigma_{s}^{2}\textbf{I}$ are the covariance matrices of the targets excluding the $i$-th, clutters, and AWGN, respectively. Then, the SCNR of the $i$-th target, $\Gamma_i$, can be calculated as (\ref{SCNR}) shown at the top of this page.
\begin{figure*}[ht] 
\centering 
\vspace*{-\baselineskip}
\begin{equation}\label{SCNR}
\begin{aligned}
 \Gamma _{i}
&=\frac{{\textstyle \sum_{j=1}^{M}} \mathbf{b}_{j}^{H} \left ( \left ( \boldsymbol{\beta}_{i} \odot \textbf{A}_{i} \right ) ^{H}\textbf{f}_{i} \textbf{f}_{i}^{H} \left ( \boldsymbol{\beta}_{i} \odot \textbf{A}_{i} \right ) \right ) \mathbf{b}_{j}  }
{ {\textstyle \sum_{j=1}^{M}}\left [ \mathbf{b}_{j}^{H}\left ( {\textstyle \sum_{u\ne i}^{N}}(\boldsymbol{\beta}_{u}\odot \mathbf{A}_{u})^{H}
\mathbf{f}_{i}\mathbf{f}_{i}^{H}
(\boldsymbol{\beta}_{u}\odot \mathbf{A}_{u}) \right ) \mathbf{b}_{j} +
 \mathbf{b}_{j}^{H} \left ( {\textstyle\sum_{c=1}^{N_c}} (\boldsymbol{\beta}_{c}\odot \mathbf{A}_{c})^{H}
\mathbf{f}_{i}\mathbf{f}_{i}^{H}
(\boldsymbol{\beta}_{c}\odot \mathbf{A}_{c}) \right ) \mathbf{b}_{j}  \right ] + \sigma_{i}^{2} \textbf{f}_{i}^{H}\textbf{f}_{i} }, \forall{i}
\end{aligned}
\end{equation}
\hrulefill 
\end{figure*}

\section{MA-aided ISAC}\label{Sec:BF design}
\subsection{Problem Formulation}
We aim to maximize the communication SSR subject to constraints on the antenna positions, SCNR of targets, and BS transmit power. The problem is formulated as
\begin{subequations}  \label{P1}
\begin{align}
  (\mathrm {P} 1  )\mathop {\max }\limits_{\left \{  \textbf{b}_{j}\right \}, \textbf{x}  }  {\text{ }}
  & SSR \label{P1a}\\
  {\text{s}}{\text{.t}}{\text{.   }}
  &\left | x_{b}-x_{a} \right |\ge d_{min},1\le a\ne b\le N_{t}, \label{P1b}\\
  &\left \{ x_{n}\right \}_{n=1}^{N_{t}}\subset  A=[L_{min},L_{max}],\label{P1c}\\
  & \sum_{j=1}^{M} \mathrm {tr}(\textbf{b}_{j}\textbf{b}_{j}^{H})\le P_{0},\label{P1d}\\
  &\Gamma_{i}\ge s_{i},\forall i\label{P1e},
\end{align}
\end{subequations}
where $d_{min}$ is the minimum distance between any two adjacent antennas, (\ref{P1c}) is the feasible moving region for all antennas, (\ref{P1d}) is the transmit power constraint, and $s_{i}$ is the SCNR constraint of the $i$-th target. Problem $(\mathrm {P} 1  )$ is non-convex due to the fractional objective (\ref{P1a}), the SCNR constraint (\ref{P1e}), and the coupling variables regarding the antenna positions. Therefore, an alternating optimization approach will be employed.

\subsection{Beamforming Design With Given \textbf{x}}
For given $ \textbf{x}$, (P1) is reformulated as
\begin{subequations}  \label{P2}
\begin{align}
  (\mathrm {P} 2  )\mathop {\max }\limits_{\left \{  \textbf{b}_{j}\right \}   }  {\text{ }}
  & SSR  \label{P2a}\\
  {\text{s}}{\text{.t}}{\text{.   }}
  & (\text{\ref{P1d}}) \text{ and } (\text{\ref{P1e}}).\label{P2b}
\end{align}
\end{subequations}
First, the Lagrangian dual transform is applied to convert the communications SR and leakage SR into \cite{10373771}
\begin{equation}
    SSR_{1} = \zeta_{c}^{+} \left ( \textbf{b},\boldsymbol{\gamma},\textbf{x}  \right )+ \zeta_{e}^{-} \left ( \textbf{b},\boldsymbol{\gamma}_{e},\textbf{x}  \right ),
\end{equation}
where $\zeta_{c}^{+} \left ( \textbf{b},\boldsymbol{\gamma},\textbf{x}  \right )$ is the communications SR and $\zeta_{e}^{-} \left ( \textbf{b},\boldsymbol{\gamma}_{e},\textbf{x}  \right )$ is the leakage SR given by, respectively
\begin{equation}\label{eq:Lagrangian comm} 
\begin{aligned}
    \zeta_{c}^{+} \left ( \textbf{b},\boldsymbol{\gamma},\textbf{x}  \right )=&\sum_{j=1}^{M}\Bigg ( \log(1+\gamma_{j})-\gamma_{j}+\\
    &\frac{(1+\gamma_{j})\textbf{b}_{j}^{H} \boldsymbol{\mathcal{H}}_{j}(\textbf{x})\textbf{b}_{j}}{\sum\limits_{k=1}^{M}  \textbf{b}^{H}_{k}\boldsymbol{\mathcal{H}}_{j}(\textbf{x})\textbf{b}_{k} +\sigma _{j}^{2}}\Bigg) ,
\end{aligned}
\end{equation}
\begin{equation} \label{eq:Lagrangian Eve}
    \begin{aligned}
        \zeta_{e}^{-} \left ( \textbf{b},\boldsymbol{\gamma}_{e},\textbf{x}  \right ) &=\sum_{j=1}^{M}\Bigg( \log(1-\gamma_{e,j})+\gamma_{e,j}-\\
        &\frac{(1-\gamma_{e,j})\textbf{b}_{j}^{H} \boldsymbol{\mathcal{H}}_{e}(\textbf{x})\textbf{b}_{j}}{{\textstyle \sum_{k=1,k\ne j}^{M}} \textbf{b}_{k}^{H} \boldsymbol{\mathcal{H}}_{e}(\textbf{x})\textbf{b}_{k} +\sigma _{e,j}^{2}}\Bigg),
    \end{aligned}
\end{equation}
where $\boldsymbol{\gamma}_{e}=[\gamma_{e,1},\dots,\gamma_{e,j},\dots,\gamma_{e,M}]$ and $\boldsymbol{\gamma}=[\gamma_{1},\dots,\gamma_{j},\dots,\gamma_{M}]$ are the auxiliary variables for the Eve and CUs, respectively, which can be derived as $\gamma_{e,j}=\frac{\textbf{b}_{j}^{H}\mathcal{H} _{e}(\textbf{x}) \textbf{b}_{j}}
{\sum\limits_{k=1}^{M}  \textbf{b}^{H}_{k}\boldsymbol{\mathcal{H}}_{e}(\textbf{x})\textbf{b}_{k} +\sigma _{e,j}^{2}}
$ and $\gamma_{j}=\frac{\textbf{b}_{j}^{H}\boldsymbol{\mathcal{H}}_{j}(\textbf{x}) \textbf{b}_{j}}
{\sum\limits_{k=1,k\ne j}^{M}  \textbf{b}^{H}_{k}\boldsymbol{\mathcal{H}}_{j}(\textbf{x})\textbf{b}_{k} +\sigma _{j}^{2}}$ \cite{8310563}. 
Then, FP is applied to transform the fractional parts in (\ref{eq:Lagrangian comm}) and (\ref{eq:Lagrangian Eve}) into 
\begin{equation}\label{eq:FP comm}
    \begin{aligned}
    &\zeta_{c}^{+} \left (\textbf{x}, \textbf{b},\boldsymbol{\gamma},\boldsymbol{\vartheta }  \right )=
    \sum_{j=1}^{M}\Bigg ( \log(1+\gamma_{j})-\gamma_{j}-\vartheta_{j}^{2}\sigma_{j}^2+\\
    & 2\vartheta_{j}\sqrt{(1+\gamma_{j})}\left | \textbf{b}_{j}^{H}\mathbf{h}_{j}(\textbf{x}) \right | -\sum_{k=1}^{M}\vartheta_{k}^{2}\textbf{b}_{k}^{H} \boldsymbol{\mathcal{H}}_{j}(\textbf{x})\textbf{b}_{k} \Bigg ),
\end{aligned}
\end{equation}
\begin{equation}\label{eq:FP Eve}
\begin{aligned}
    &\zeta_{e}^{-} \left ( \textbf{x},\textbf{b},\boldsymbol{\gamma}_{e},\boldsymbol{\vartheta }_{e}  \right )=\sum_{j=1}^{M}\Bigg ( 
     \log(1-\gamma_{e,j})+\gamma_{e,j}+\vartheta_{e,j}^{2}\sigma_{e,j}^2\\
     &-2\vartheta_{e,j}\sqrt{(1-\gamma_{e,j}) }\left | \textbf{b}_{j}^{H}\mathbf{h} _{e}(\textbf{x}) \right | +\sum_{p=1,p\ne j}^{M}\vartheta_{e,p}^{2}\textbf{b}_{p}^{H}\mathcal{H} _{e}(\textbf{x}) \textbf{b}_{p}\Bigg )  ,
\end{aligned}
\end{equation}
where $ \boldsymbol{\vartheta }=[\vartheta_{1},\dots,\vartheta_{j},\dots,\vartheta_{M}]$ and $ \boldsymbol{\vartheta }_{e}=[\vartheta_{e,1},\dots,\vartheta_{e,j},\dots,\vartheta_{e,M}]$ are the auxiliary variables for the fractional parts of (\ref{eq:FP comm}) and (\ref{eq:FP Eve}), respectively, which can derived as $\vartheta_{e,j}=\frac{\sqrt{(1-\gamma_{e,j})\textbf{b}_{j}^{H}\mathcal{H} _{e}(\textbf{x}) \textbf{b}_{j}}}
{\sum_{p=1,p\ne j}^{M}\textbf{b}_{p}^{H}\mathcal{H} _{e}(\textbf{x}) \textbf{b}_{p}+\sigma_{e,p}^{2}} $ and $\vartheta_{j}=\frac{\sqrt{(1+\gamma_{j})\textbf{b}_{j}^{H}\boldsymbol{\mathcal{H}}_{j}(\textbf{x}) \textbf{b}_{j}}}
{\sum_{k=1}^{M}\textbf{b}_{k}^{H}\boldsymbol{\mathcal{H}}_{j} (\textbf{x})\textbf{b}_{k}+\sigma_{j}^{2}}$ \cite{8310563}.
Notice that $\zeta_{c} \left (\textbf{x}, \textbf{b},\boldsymbol{\gamma},\boldsymbol{\vartheta }  \right )$ and $\zeta_{e}^{-} \left ( \textbf{x},\textbf{b},\boldsymbol{\gamma}_{e,j},\boldsymbol{\vartheta }_{e,j}  \right )$ are neither concave nor convex. Thus, the SSR is still non-convex.
For any convex function $f(\textbf{b}_{j} )$, by applying the first-order convex approximation, we have
\begin{equation}
    f(\textbf{b}_{j} )\ge f(\textbf{b}_{j0} )+\Re \left \{ \nabla ^H f(\textbf{b}_{j0} )[\textbf{b}_{j}-\textbf{b}_{j0}]\right \} ,
\end{equation}
where $\textbf{b}_{j0}$ can be any feasible point. Then, we use the convex approximation to approximate the convex part in (\ref{eq:FP comm}) as
\begin{equation}
    \begin{aligned}\label{appro for rate}
    &2\vartheta_{j}\sqrt{(1+\gamma_{j})}\left | \textbf{b}_{j}^{H}\mathbf{h} _{j}(\textbf{x}) \right |\ge 
    2\vartheta_{j}\sqrt{(1+\gamma_{j})\textbf{b}_{j0}^{H}\boldsymbol{\mathcal{H}}_{j}(\textbf{x}) \textbf{b}_{j0}}+\\
    &2\vartheta_{j}\sqrt{1+\gamma_{j}}\frac{\Re\left \{ \textbf{b}_{j0}^{H}\boldsymbol{\mathcal{H}}_{j}(\textbf{x})[\textbf{b}_{j}-\textbf{b}_{j0}] \right \} }{\sqrt{\textbf{b}_{j0}^{H}\boldsymbol{\mathcal{H}}_{j}(\textbf{x}) \textbf{b}_{j0}} }, \forall j.
    \end{aligned}
\end{equation}
The lower bound of the communication sum rate can be obtained as (\ref{eq:Concave comm}) on the top of this page.
\begin{figure*}[!t] 
\centering 
\vspace*{-\baselineskip}
\begin{equation}\label{eq:Concave comm}
    \begin{aligned}
    &\Theta _{c}^{+} \left (\textbf{x}, \textbf{b},\boldsymbol{\gamma},\boldsymbol{\vartheta }  \right )
    =\\
    &\sum_{j=1}^{M}\Bigg ( \log(1+\gamma_{j})-\gamma_{j}-\vartheta_{j}^{2}\sigma_{j}^2+
     2\vartheta_{j}\sqrt{(1+\gamma_{j})}\left | \textbf{b}_{j0}^{H}\mathbf{h}_{j}(\textbf{x}) \right |+
2\vartheta_{j}\sqrt{1+\gamma_{j}}\frac{\Re\left \{ \textbf{b}_{j0}^{H}\boldsymbol{\mathcal{H}}_{j}(\textbf{x})[\textbf{b}_{j}-\textbf{b}_{j0}] \right \} }{\sqrt{\textbf{b}_{j0}^{H}\boldsymbol{\mathcal{H}}_{j}(\textbf{x}) \textbf{b}_{j0}} } -\sum_{k=1}^{M}\vartheta_{k}^{2}\textbf{b}_{k}^{H} \boldsymbol{\mathcal{H}}_{j}(\textbf{x})\textbf{b}_{k} \Bigg )
\end{aligned}
\end{equation} 
\hrulefill 
\end{figure*}
Similarly, the convex part in (\ref{eq:FP Eve}) is approximated as
\begin{equation}
    \begin{aligned}
        \textbf{b}_{p}^{H}\boldsymbol{\mathcal{H}} _{e}(\textbf{x}) \textbf{b}_{p}&\ge \textbf{b}_{p0}^{H}\boldsymbol{\mathcal{H}} _{e}(\textbf{x}) \textbf{b}_{p0}
+2\Re \left \{ \textbf{b}_{p0}^{H}\boldsymbol{\mathcal{H}} _{e}(\textbf{x})\left ( \textbf{b}_{p}-\textbf{b}_{p0} \right )  \right \} \\
&=\Re \left \{ 2\textbf{b}_{p0}^{H}\boldsymbol{\mathcal{H}} _{e}(\textbf{x})\textbf{b}_{p} - \textbf{b}_{p0}^{H}\boldsymbol{\mathcal{H}} _{e}(\textbf{x})\textbf{b}_{p0}  \right \}.
    \end{aligned}
\end{equation}
Then, the upper bound of the leakage SR can be obtained as (\ref{Concave Eve}) on the top of this page.
\begin{figure*}[!t] 
\centering 
\vspace*{-\baselineskip}
\begin{equation}\label{Concave Eve}
\begin{aligned}
    &\Theta_{e}^{-} \left ( \textbf{x},\textbf{b},\boldsymbol{\gamma}_{e},\boldsymbol{\vartheta }_{e}  \right )
    =\\ 
    &\sum_{j=1}^{M}\Bigg ( 
     \log(1-\gamma_{e,j})+\gamma_{e,j}+\vartheta_{e,j}^{2}\sigma_{e,j}^2
     -2\vartheta_{e,j}\sqrt{(1-\gamma_{e,j}) }\left | \textbf{b}_{j}^{H}\mathbf{h} _{e}(\textbf{x}) \right | +\sum_{p=1,p\ne j}^{M}\vartheta_{e,p}^{2}\Re \left \{ 2\textbf{b}_{p0}^{H}\boldsymbol{\mathcal{H}} _{e}(\textbf{x})\textbf{b}_{p} -\textbf{b}_{p0}^{H}\boldsymbol{\mathcal{H}} _{e}(\textbf{x})\textbf{b}_{p0} \right \}  \Bigg )  
\end{aligned}
\end{equation}
\hrulefill
\end{figure*}
Next, the lower bound of the SSR can be derived as
\begin{equation}
    SSR_{2} = \Theta _{c}^{+} \left (\textbf{x}, \textbf{b},\boldsymbol{\gamma},\boldsymbol{\vartheta }  \right )+\Theta _{e}^{-} \left (\textbf{x}, \textbf{b},\boldsymbol{\gamma}_e,\boldsymbol{\vartheta }_e  \right ).
\end{equation}

For the fractional SCNR constraint (\ref{P1e}), we first define $\Xi_{i}^{}=\left ( \boldsymbol{\beta}_{i} \odot \textbf{A}_{i} \right ) ^{H}\textbf{f}_{i} \textbf{f}_{i}^{H} \left ( \boldsymbol{\beta}_{i} \odot \textbf{A}_{i} \right )$, $\Xi_{u,i}^{}={\textstyle \sum_{u\ne i}^{N}}(\boldsymbol{\beta}_{u}\odot \mathbf{A}_{u})^{H}\mathbf{f}_{i}\mathbf{f}_{i}^{H}(\boldsymbol{\beta}_{u}\odot \mathbf{A}_{u})$, and $\Xi_{c,i}^{}={\textstyle\sum_{c=1}^{N_c}} (\boldsymbol{\beta}_{c}\odot \mathbf{A}_{c})^{H}\mathbf{f}_{i}\mathbf{f}_{i}^{H}(\boldsymbol{\beta}_{c}\odot \mathbf{A}_{c}) $. Next, (\ref{P1e}) can be written as
\begin{equation}
    \begin{aligned}
&{\textstyle \sum_{j=1}^{M}}\mathbf{b}_{j}^{H}\Xi_{u,i}^{}  \mathbf{b}_{j} + {\textstyle \sum_{j=1}^{M}}\mathbf{b}_{j}^{H} \Xi_{c,i}^{}  \mathbf{b}_{j}+\\
&\sigma_{i}^{2} \textbf{f}_{i}^{H}\textbf{f}_{i}-
\frac{ {\textstyle \sum_{j=1}^{M}}\textbf{b}_{j}^{H}\Xi _{i}^{}\textbf{b}_{j} }{s_i} \le 0,\forall i .
\end{aligned}
\end{equation}
By applying the first-order convex approximation, the concave parts of (\ref{P1e}) are converted into
\begin{equation}
    -\textbf{b}_{j}^{H}\Xi _{i}^{} \textbf{b}_{j}\le -\textbf{b}_{j0}^{H}\Xi _{i}^{} \textbf{b}_{j0}
        -2\Re \left \{ \textbf{b}_{j0}^{H}\Xi _{i}^{}\left ( \textbf{b}_{j}-\textbf{b}_{j0} \right )  \right \}, \forall i.
\end{equation}
Then, (\ref{P1e}) can be rewritten as (\ref{Convex SCNR}) on the top of this page.
\begin{figure*}
    \centering
    \vspace*{-\baselineskip}
\begin{equation}\label{Convex SCNR}
         \sum_{j=1}^{M}\left ( \mathbf{b}_{j}^{H}\Xi_{u,i}^{}  \mathbf{b}_{j} +\mathbf{b}_{j}^{H} \Xi_{c,i}^{}  \mathbf{b}_{j} \right )  + \sigma_{i}^{2} \textbf{f}_{i}^{H}\textbf{f}_{i}+
        \frac{ {\textstyle \sum_{j=1}^{M}} \Re \left \{ \textbf{b}_{j0}^{H}\Xi _{i}^{}\textbf{b}_{j0}-2\textbf{b}_{j0}^{H}\Xi _{i}^{}\textbf{b}_{j}  \right \} }{s_i} \le 0,\forall i .
    \end{equation}
    \hrulefill 
\end{figure*}
Finally, (P2) can be reformulated as
\begin{subequations}  \label{P2.1}
\begin{align}
  (\mathrm {P} 2.1 )\mathop {\max }\limits_{\left \{  \textbf{b}_{j}\right \}   }  {\text{ }}
  & SSR_{2}  \label{P2.1a}\\
  {\text{s}}{\text{.t}}{\text{.   }}
  &(\text{\ref{P1d}}) \text{ and } (\text{\ref{Convex SCNR}}),\label{P2.1b}
\end{align}
\end{subequations}
which can be solved by CVX to obtain a suboptimal solution \cite{grant2014cvx}.
The procedure is outlined in Algorithm \ref{algorithm 1}. The computational complexity is $\mathcal{O}(K(MN_{t})^3)$ for $K$ iterations. Since the optimized SSR is upper bounded and increases monotonically with each iteration, convergence is guaranteed.

\begin{algorithm}[]
\caption{Proposed Iterative Algorithm for Beamforming Design in (\ref{P2})}\label{algorithm 1}
\footnotesize
\begin{algorithmic}[1]
\STATE \textbf{Input:} Convergence threshold $\epsilon$; iteration counter $ite=1$; initial beamformers $\{\textbf{b}_{j(ite)}\}$ initialized by the matched-filter (maximum ratio transmission), with the total transmit power $P_0$ split equally across the $M$ beams, i.e., $\mathbf{b}_{j(ite)} = \sqrt{\frac{P_0}{M}}\frac{\mathbf{h}_{j}(\textbf{x})}{\left \| \mathbf{h}_{j}(\textbf{x}) \right \|}\in\mathbb{C}^{N_t\times 1}$; the initial receive beamformers $\{\textbf{f}_{i(ite)}\}$ obtained by $\{\textbf{b}_{j(ite)}\}$; the initial auxiliary variables $\boldsymbol{\gamma}^{ite}$,$\boldsymbol{\vartheta }^{ite}$, $\boldsymbol{\gamma}_{e}^{ite}$, and $\boldsymbol{\vartheta }_{e}^{ite}$ obtained by $\{\textbf{b}_{j(ite)}\}$; the initial SSR $SSR^{(ite)}$ obtained by $\{\textbf{b}_{j(ite)}\}$; the antenna positions $\textbf{x}$.
\STATE \textbf{Output:} Optimized beamformers $\{\textbf{b}_{j}\}$.
\STATE Generate channels for CUs and steering/focusing vectors for targets using the given antenna positions $\textbf{x}$.
\REPEAT
    \STATE Solve (\ref{P2.1}) to obtain $\{\textbf{b}_{j}\}$.
    \STATE Update $ite = ite + 1$.
    \STATE Update $\{\textbf{b}_{j(ite)}\}$, $\boldsymbol{\gamma}^{ite}$,$\boldsymbol{\vartheta }^{ite}$, $\boldsymbol{\gamma}_{e}^{ite}$,$\boldsymbol{\vartheta }_{e}^{ite}$, $\{\textbf{f}_{i(ite)}\}$, and $SSR^{(ite)}$.
\UNTIL{$\left | SSR^{(ite)} - SSR^{(ite-1)} \right | \leq \epsilon$}
\end{algorithmic}
\end{algorithm}


\subsection{Antenna Position Optimization With Given Beamformer}
For given beamformer $\{\textbf{b}_{j}\}$, the antenna positions $\textbf{x}$ can be updated by solving
\begin{subequations}  \label{P3}
\begin{align}
  (\mathrm {P} 3  )\mathop {\max }\limits_{\textbf{x} }  {\text{ }}
  & \zeta_{c}^{+} \left (\textbf{x}, \textbf{b},\boldsymbol{\gamma},\boldsymbol{\vartheta }  \right )+\zeta_{e}^{-} \left ( \textbf{x},\textbf{b},\boldsymbol{\gamma}_{e},\boldsymbol{\vartheta }_{e}  \right ) \textcolor{red}{\label{P3a}}\\
  {\text{s}}{\text{.t}}{\text{.   }}
  & (\text{\ref{P1b}} )  \text{ and } (\text{\ref{P1c}}),
\end{align}
\end{subequations}
where (\ref{P1e}) is omitted since it does not influence the antenna position update for secrecy rate maximization. The objective function $\zeta_{c} \left (\textbf{x}, \textbf{b},\mathbf{\gamma},\mathbf{\vartheta } \right )$ is differentiable with respect to $\textbf{x}$ but is neither convex nor concave. Therefore, the projected gradient ascent (PGA) method can be employed to identify locally optimal antenna positions \cite{10839251}, \cite{10164189}. The detailed procedure is as follows.

\subsubsection{\textbf{Initialize antenna positions}} 
To generate feasible initial antenna positions satisfying the range
constraint (\ref{P1b}) and the minimum inter-antenna separation constraint (\ref{P1c}), a one-dimensional grid search over the uniform inter-antenna spacing is performed.
Specifically, the candidate spacing set is defined as
\begin{equation}\label{eq:spacing set}
    \mathcal{D} = \left\{ d=d_{\min} + q\frac{\lambda}{4} \ \Big|\ q = 0, 1, \dots,\left\lfloor
     \frac{A - (N_t - 1)d_{\min}}{(N_t - 1){\lambda}/{4}} \right\rfloor \right\},
\end{equation}
where $A = L_{\max} - L_{\min}$ is the maximum allowed array aperture, $d$ is the spacing between any adjacent antennas. For each candidate spacing
$d \in \mathcal{D}$, a uniformly spaced antenna array is constructed as
\begin{equation}
    x_n(d) = L_{\min} + (n-1)d, \quad n = 1, \dots, N_t,
\end{equation}
which guarantees that all antenna positions lie within $[L_{\min}, L_{\max}]$ and that
the separation between every pair of adjacent antennas equals $d \geq d_{\min}$, thereby
satisfying constraints (\ref{P1b}) and (\ref{P1c}) by construction. The SSR of each candidate configuration is evaluated via Algorithm \ref{algorithm 1}, and the optimal spacing is selected as
\begin{equation}
    d^{*} = \underset{d\, \in\, \mathcal{D}}{\arg\max}\ SSR_2.
\end{equation}
The initial antenna positions are then set as $x_n^{(0)} = L_{\min} + (n-1)d^{*}$,
$n = 1, \dots, N_t$. This one-dimensional grid search is guaranteed to provide a feasible and effective starting point.

\subsubsection{\textbf{Apply correct propagation models}} Based on the aperture $D = x_{N_t} - x_1$, compute $\xi_j$ and $\omega_j$ for every CU, target, clutter, and Eve using (\ref{eq:omega}), then form their responses via (\ref{eq:unified response}). Devices with $\omega_j>0$ are assigned to the NF sets ($M_1$, $N_1$), while those with $\omega_j=0$ are assigned to the FF sets ($M_2$, $N_2$). Devices within the transition window simply use the $\omega_j$-weighted response. This step is needed only for MA, whose aperture varies during optimization.
Compared with a binary Rayleigh-distance rule, (\ref{eq:omega}) offers the advantage that $\Lambda(\mathbf{x})$ in Step 3) remains continuous in $\mathbf{x}$ across the NF/FF boundary, preventing abrupt model switches between position updates. The transition window is set as $[\xi_{lo},\xi_{hi}]=[0.95,1.05]$.

\subsubsection{\textbf{Calculate gradient ascent}}
With fixed $\textbf{b},\mathbf{\gamma},\mathbf{\vartheta }$, we denote \begin{equation} \label{eq:Lambda}
    \Lambda \left (\textbf{x}  \right )=\zeta_{c}^{+} \left (\textbf{x}, \textbf{b},\boldsymbol{\gamma},\boldsymbol{\vartheta }  \right )+\zeta_{e}^{-} \left ( \textbf{x},\textbf{b},\boldsymbol{\gamma}_{e},\boldsymbol{\vartheta }_{e}  \right ).
\end{equation}
The gradient of $\Lambda \left (\textbf{x}  \right )$ with respect to $\textbf{x}=\left \{ x_{n} \right \}_{n=1}^{N_{t}}$ is given by
\begin{equation}\label{deri Theta}
    \bigtriangledown_{\textbf{x}}\Lambda \left (\textbf{x}  \right )=
\left [ \frac{\partial \Lambda \left (\textbf{x}  \right ) }{\partial x_{1}},\frac{\partial \Lambda \left (\textbf{x}  \right )}{\partial x_{2}},\dots, \frac{\partial \Lambda \left (\textbf{x}  \right )}{\partial x_{N_{t}}} \right ] ^{T}.
\end{equation}

The full derivation is referred to Appendix \ref{app:gradient}, and the resulting gradient is given by (\ref{eq:Final gradient}) at the top of the next page, where $\bigtriangledown_{\mathbf{x}}\mathcal{G}_{j,k}$, $\bigtriangledown_{\mathbf{x}}\mathcal{G}_{e,p}$, $\bigtriangledown_{\mathbf{x}}\mathcal{F}_j$, and $\bigtriangledown_{\mathbf{x}}\mathcal{F}_{ej}$ are the closed-form per-antenna gradients.
\begin{figure*}
    \centering
    \vspace*{-\baselineskip}
\begin{equation}\label{eq:Final gradient}
    \begin{aligned}
        \frac{\partial \Lambda \left (\textbf{x}  \right ) }{\partial \textbf{x}}=\sum_{j=1}^{M}\left ( 2\vartheta_{j}\sqrt{(1+\gamma_{j})} \frac{\bigtriangledown_{\textbf{x}}\mathcal{G}_{j,j}}{2\sqrt{\textbf{b}_{j}^{H} \mathcal{H}_{j}(\textbf{x})\textbf{b}_{j}} } -\sum_{k=1}^{M}\vartheta_{k}^{2}\bigtriangledown_{\textbf{x}} \mathcal{G}_{j,k}-2\vartheta_{e,j}\sqrt{(1-\gamma_{e,j})}\frac{\bigtriangledown_{\textbf{x}}\mathcal{G}_{e,j}}{2\sqrt{\textbf{b}_{j}^{H} \mathcal{H}_{e}(\textbf{x})\textbf{b}_{j}} }+  \sum_{p=1,p\ne j}^{M}\vartheta_{e,p}^{2}\bigtriangledown_{\textbf{x}} \mathcal{G}_{e,p}   \right )
    \end{aligned}
\end{equation}
\hrulefill
\end{figure*}

Using the unified response (\ref{eq:unified response}), the gradients in (\ref{eq:Final gradient}) reduce to a single $\omega_j$-weighted expression, with the pure NF ($\omega_j=1$) and FF ($\omega_j=0$) cases in Appendix \ref{app:gradient} as special cases. This makes $\Lambda(\mathbf{x})$ continuously differentiable across the NF/FF boundary. Each antenna position $x_n$ affects the objective in two ways: directly through the array response, and indirectly through the weights $\{\omega_j\}$, which depend on the aperture $D=x_{N_t}-x_1$. Applying the chain rule, the total derivative is
\begin{equation}\label{eq:total gradient}
\frac{\partial\Lambda(\mathbf{x})}{\partial x_n}=\left.\frac{\partial\Lambda(\mathbf{x})}{\partial x_n}\right|_{\boldsymbol{\omega}}+\sum_j\frac{\partial\Lambda(\mathbf{x})}{\partial\omega_j}\frac{\partial\omega_j}{\partial D}\frac{\partial D}{\partial x_n},
\end{equation}
where the first term is the gradient from (\ref{eq:Final gradient}) with $\{\omega_j\}$ fixed, and the second term accounts for how changes in the aperture $D$ shift the NF/FF weights. Using $\partial\omega_j/\partial D$ from (\ref{eq:omega}), this aperture term becomes
\begin{equation}\label{eq:aperture gradient}
\begin{aligned}
&\left(\frac{\partial\Lambda(\mathbf{x})}{\partial x_n}\right)_{\mathrm{ap}}
=\frac{\partial D}{\partial x_n}\times\\
&\sum_{j}
\frac{\partial\Lambda(\mathbf{x})}{\partial\omega_j}
\frac{4D\sin^2(\theta_j)}
{\lambda r_j(\xi_{hi}-\xi_{lo})}
S'\!\left(\frac{\xi_j-\xi_{lo}}{\xi_{hi}-\xi_{lo}}\right),
\end{aligned}
\end{equation}
where
\begin{equation}\label{eq:dDdx}
\frac{\partial D}{\partial x_n}=
\begin{cases}
+1, & n=N_t,\\
-1, & n=1,\\
0, & \text{otherwise}.
\end{cases}
\end{equation}

\subsubsection{\textbf{Line search and feasibility check}}
For a fixed beamformer, Steps 2)-5) are repeated as an inner loop until the projected-gradient residual satisfies $\|\mathbf{x}-\mathfrak{D}\{\mathbf{x}+\eta_0\mathbf{d}\}\|\le\epsilon_x$. The beamformer is then updated by Algorithm \ref{algorithm 1}. A fixed step may cause oscillations; we therefore adopt a backtracking (Armijo) line search, which also performs the sensing feasibility check since $\mathbf{x}$ enters $\Gamma_i$ in (\ref{P1e}), whose trial step follows the Barzilai-Borwein (BB) rule \cite{barzilai1988two}: with $\mathbf{d}=\bigtriangledown_{\mathbf{x}}\Lambda(\mathbf{x})$,
\begin{equation}\label{eq:bb step}
    \eta_0=\min\left\{\eta_{BB}, \frac{\lambda/4}{\|\mathbf{d}\|_\infty}\right\},\quad
    \eta_{BB}=\frac{\mathbf{s}^T\mathbf{s}}{-\mathbf{s}^T\mathbf{y}},
\end{equation}
where $\mathbf{s}$ and $\mathbf{y}$ track the changes in $\mathbf{x}$ and $\mathbf{d}$ over the last two inner iterations with a fixed beamformer. At the first inner iteration after each beamformer update, and whenever $\mathbf{s}^T\mathbf{y}\ge0$, reset $\eta_{BB}$ to $c_0d_{\min}/\|\mathbf{d}\|_\infty$. The $\lambda/4$ cap limits each antenna to a quarter-wavelength move per trial. With backtracking factor $\rho\in(0,1)$ and Armijo constant $c\in(0,1)$, pick the largest step $\eta=\eta_0\rho^k$, $k\le k_{\max}$, such that the projected candidate $\tilde{\mathbf{x}}=\mathfrak{D}\{\mathbf{x}+\eta\mathbf{d}\}$ satisfies the ascent condition along the projection arc \cite{bertsekas1999nonlinear}
\begin{equation}\label{eq:armijo}
    \Lambda(\tilde{\mathbf{x}}) \ge \Lambda(\mathbf{x}) + c\,\mathbf{d}^T\left(\tilde{\mathbf{x}}-\mathbf{x}\right)
\end{equation}
and the feasibility condition that the left-hand side of (\ref{Convex SCNR}) evaluated at the candidate $\tilde{\mathbf{x}}$ with the current beamformer remains within a tolerance $\delta_\Gamma>0$ for all targets. If no such $\eta$ exists, the update is rejected.

\subsubsection{\textbf{Feasible range projection}}
The feasible set is the convex polyhedron $\mathcal{X}=\{\mathbf{x}: x_1\ge L_{\min},\ x_{n}\ge x_{n-1}+d_{\min},\ n=2,\dots,N_t,\ x_{N_t}\le L_{\max}\}$, and the position update is the projected gradient step
\begin{equation}\label{eq:update rule}
    \mathbf{x}\leftarrow\mathfrak{D}\left\{\mathbf{x}+\eta\,\mathbf{d}\right\},\quad
    \mathfrak{D}\{\mathbf{z}\}=\arg\min_{\mathbf{x}'\in\mathcal{X}}\left\|\mathbf{x}'-\mathbf{z}\right\|^2 .
\end{equation}
This projection is computed in $\mathcal{O}(N_t)$ as follows. Define $y_n=x_n-(n-1)d_{\min}$, then the spacing constraints become $y_1\le\dots\le y_{N_t}$ and the range constraints become the box $L_{\min}\le y_n\le L_{\max}-(N_t-1)d_{\min}$. The projection reduces to isotonic regression, solved by the pool-adjacent-violators algorithm (PAVA) \cite{best1990active,deLeeuw2009isotone}: (i) $\tilde{y}_n=z_n-(n-1)d_{\min}$, (ii) $\hat{\mathbf{y}}=\mathrm{PAVA}(\tilde{\mathbf{y}})$ (merges adjacent order violations into block averages in one pass), (iii) clip $\hat{y}_n$ to $[L_{\min},L_{\max}-(N_t-1)d_{\min}]$, (iv) $x_n=\hat{y}_n+(n-1)d_{\min}$. This returns the feasible point closest to the gradient step, preserving the ascent direction.

The overall iterative algorithm for solving (P1) is summarized in Algorithm \ref{algorithm 2}. The convergence performance of the Algorithm \ref{algorithm 2} is shown in Fig. \ref{fig:convergence} in Section \ref{Sec:results}.
\begin{algorithm}[]
\caption{Iterative Algorithm for Solving (P1)}\label{algorithm 2}
\footnotesize
\begin{algorithmic}[1]
\STATE Set $ite_2=0$, $\epsilon_x$, $c_0$, $\rho$, $c$, $k_{\max}$, and $\delta_\Gamma$; initialize $\{\textbf{b}_{j}\}^{ite_2}$ by the matched-filter.
\STATE Initialize antenna positions $\textbf{x}$ according to Step 1) in Section III.C.
\REPEAT
    \STATE Set $ite_1=0$ and reset the BB pair $(\mathbf{s},\mathbf{y})$.
    \REPEAT[inner loop, beamformer $\{\textbf{b}_{j}\}^{ite_2}$ fixed]
        \STATE Update $D$, the weights $\{\omega_j\}$ by (\ref{eq:omega}), and the responses by (\ref{eq:unified response}) (Step 2)).
        \STATE Calculate $\mathbf{d}=\bigtriangledown_{\textbf{x}}\Lambda(\textbf{x})$ (Step 3)).
        \STATE Set $\eta_0$ by (\ref{eq:bb step}). Find the largest $\eta=\eta_0\rho^k$ and $k\le k_{\max}$ that satisfies (\ref{eq:armijo}) and the feasibility check (Step 4)). If none exists, keep $\textbf{x}$.
        \STATE Update $\textbf{x}\leftarrow\mathfrak{D}\{\textbf{x}+\eta\mathbf{d}\}$ by (\ref{eq:update rule}) (Step 5)); $ite_1\leftarrow ite_1+1$.
    \UNTIL{$\|\mathbf{x}-\mathfrak{D}\{\mathbf{x}+\eta_0\mathbf{d}\}\|\le\epsilon_x$ or the step is rejected.}
    \STATE Update $\{\textbf{b}_{j}\}^{ite_2+1}$ by Algorithm \ref{algorithm 1} with the updated $\textbf{x}$.
    \STATE Keep $(\textbf{x},\{\textbf{b}_{j}\})$ if its SSR is the highest.
    \STATE Update $ite_2 = ite_2+1$.
\UNTIL{The SSR converges to the desired accuracy.}
\end{algorithmic}
\end{algorithm}

\subsection{Complexity Analysis}
Table \ref{tab:complexity} summarizes the per-stage complexity. Solving (P2.1) costs $\mathcal{O}(K_1(MN_t)^3)$, where $K_1$ is the number of FP/SCA iterations, and the gradient (\ref{eq:Final gradient}) costs $\mathcal{O}(M^2N_t^2)$. This gradient cost is the same for the NF and FF parts, and the unified response (\ref{eq:unified response}) adds $\mathcal{O}(N_t)$ per link plus the $\mathcal{O}(M^2N_t)$ aperture term (\ref{eq:aperture gradient}). The line search adds a factor $\mathcal{O}(k_{\max})$ per inner iteration, the PAVA projection $\mathcal{O}(N_t)$ per step, and the one-time grid-search initialization $\mathcal{O}(|\mathcal{D}|K_1(MN_t)^3)$ with $|\mathcal{D}|=\lfloor\frac{A-(N_t-1)d_{min}}{(N_t-1)\lambda/4}\rfloor+1$. With $I_o$ and $I_i$ the numbers of outer and inner iterations, the total complexity of Algorithm \ref{algorithm 2} is $\mathcal{O}\big(|\mathcal{D}|K_1(MN_t)^3+I_o(I_i k_{\max}M^2N_t^2+K_1(MN_t)^3)\big)$, which is dominated by the beamforming term.
\begin{table}[t]
\centering
\caption{Per-stage computational complexity}
\label{tab:complexity}
\renewcommand{\arraystretch}{1.05}
\begin{tabular}{p{3.6cm}p{4.2cm}}
\toprule
\textbf{Stage} & \textbf{Complexity} \\
\midrule
Beamforming (Alg. \ref{algorithm 1}, per call) & $\mathcal{O}(K_1(MN_t)^3)$ \\
Position gradient (Step 3) & $\mathcal{O}(M^2N_t^2)$ -- same for NF/FF \\
Line search + feasibility check & $\mathcal{O}(k_{\max}(M^2N_t^2+NMN_t^2))$ \\
Grid-search initialization & $\mathcal{O}(|\mathcal{D}|K_1(MN_t)^3)$ \\
\midrule
\textbf{Algorithm \ref{algorithm 2} (total)} & $\mathcal{O}\big(|\mathcal{D}|K_1(MN_t)^3+I_o(I_i k_{\max}M^2N_t^2+K_1(MN_t)^3)\big)$ \\
\bottomrule
\end{tabular}
\end{table}

\section{Numerical Results And Discussion}\label{Sec:results}
In the numerical examples, a mixed-field ISAC system is considered, where the BS operates at $f=28$ GHz and is equipped with $N_{t}=N_{r}=Q$ antennas. This frequency has practical significance, as sensing-assisted communication can improve beam tracking performance by enabling the acquisition of user position information for optimization \cite{9171304}. The minimum distance between adjacent antennas is set to $d_{min}=\frac{\lambda}{2}$. Unless otherwise specified, the Rician factor is set to $\mathcal{K}=3$ with $L=1$ NLoS path. 
The angular offset standard deviation is set to $\sigma_\theta=2^{\circ }$, i.e., $\Delta\theta \sim \mathcal{N}(0, (2^\circ)^2)$, and the distance offset range is set to $\Delta r_{\max}=1$ m, i.e., $\Delta r \sim \mathcal{U}(0, 1)$.
All numerical results are obtained using $200$ Monte-Carlo simulations.
The total transmit power is set to $30$ dBm, and the noise power is set to $-90$ dBm. 
The line-search parameters are $c_0=0.25$, $\rho=0.5$, $c=10^{-4}$, $k_{\max}=15$, with the displacement cap $\lambda/4$ in (\ref{eq:bb step}). Set $\epsilon_x=10^{-3}\lambda$, the transition window in (\ref{eq:omega}) is $[\xi_{lo},\xi_{hi}]=[0.95,1.05]$, and the number of samples is $T=100$.
For performance comparison, six schemes are considered to examine the effect of propagation model change and the difference between MA and FA: 
\begin{enumerate}
    \item FA-FF: FA-ISAC that always uses the FF model even in mixed-field.
    \item FA-NF: FA-ISAC that always uses the NF model even in mixed-field. This requires NF channel estimation for all devices and hence, has the highest complexity.
    \item FA-mix: FA-ISAC that uses the actual mixed-field models, that is, NF models for NF CUs/targets and FF models for FF CUs/targets.
    \item MA-FF: Conventional MA-ISAC that always uses the FF model and disregards the propagation model changes in antenna position updates, as in \cite{10839251}, as CSI is acquired before optimization is performed.
    \item MA-NF: MA-ISAC that always uses the NF models, as in \cite{11075964}, as FF is an approximation of NF but this has the highest complexity requiring NF channel estimation for all devices. This gives the performance upper limit of MA-ISAC.
    \item MA-mix: The proposed MA-ISAC that uses the actual mixed-field models considering the propagation model changes in antenna position updates.
\end{enumerate}
Using the unified response in (\ref{eq:unified response}), FF corresponds to $\omega_j\equiv0$ and NF corresponds to $\omega_j\equiv1$ for all devices, while mixed-field schemes compute $\omega_j$ via (\ref{eq:omega}). All schemes are evaluated against channels generated from the correct mixed-field model.
\begin{table*}[!t]
\centering
\caption{Computational complexity and measured runtime comparison of the proposed and benchmark algorithms.}
\label{tab:runtime}
\begin{tabular}{lccc}
\toprule
\textbf{Algorithm} & \textbf{Computational complexity} & \textbf{Runtime (s)} & \textbf{Achieved SSR (bps/Hz)} \\
\midrule
Proposed (Algorithm~2) &
$\mathcal{O}\!\left( I_{\mathrm{o}} \left[ I_{1} \left( N_t \bar{K} \right)^{3.5} + I_{\mathrm{i}} k_{\mathrm{A}} N_t^{2} \bar{J} \right] \right)$ &
$166.6$ & $20.06$ \\
PSO-based benchmark &
$\mathcal{O}\!\left( N_{\mathrm{p}} T_{\mathrm{p}} I_{1} \left( N_t \bar{K} \right)^{3.5} \right)$ &
$291.1$ & $20.56$ \\
BnB-based benchmark &
$\mathcal{O}\!\left( N_{\mathrm{ev}} I_{1} \left( N_t \bar{K} \right)^{3.5} \right)$ &
$48.7$ & $19.43$ \\
\bottomrule
\end{tabular}
\vspace{2pt}
\begin{flushleft}
\footnotesize
$I_{1}$: SCA iterations of Algorithm~1 per beamforming solve; $\bar{K}=M+N$: number of communication and sensing beams; $\bar{J}$: total number of CUs, Eve, targets, and clutters; $I_{\mathrm{o}}$/$I_{\mathrm{i}}$: outer/inner iterations of Algorithm~2; $k_{\mathrm{A}}$: Armijo backtracking trials per inner step; $N_{\mathrm{p}}$, $T_{\mathrm{p}}$: PSO swarm size and iteration count ($N_{\mathrm{p}}=6$, $T_{\mathrm{p}}=8$); $N_{\mathrm{ev}}$: number of candidate uniform spacings evaluated by the BnB search, each requiring a full beamforming solve ($N_{\mathrm{ev}}=8$).
\end{flushleft}
\end{table*}

\subsection{Effect of $A$, $P_0$, and $Q$}
\begin{figure*}[!t]
    \centering
    \captionsetup{font={small}}
    \begin{minipage}[t]{0.3\textwidth}
        \centering
        \includegraphics[width=\linewidth]{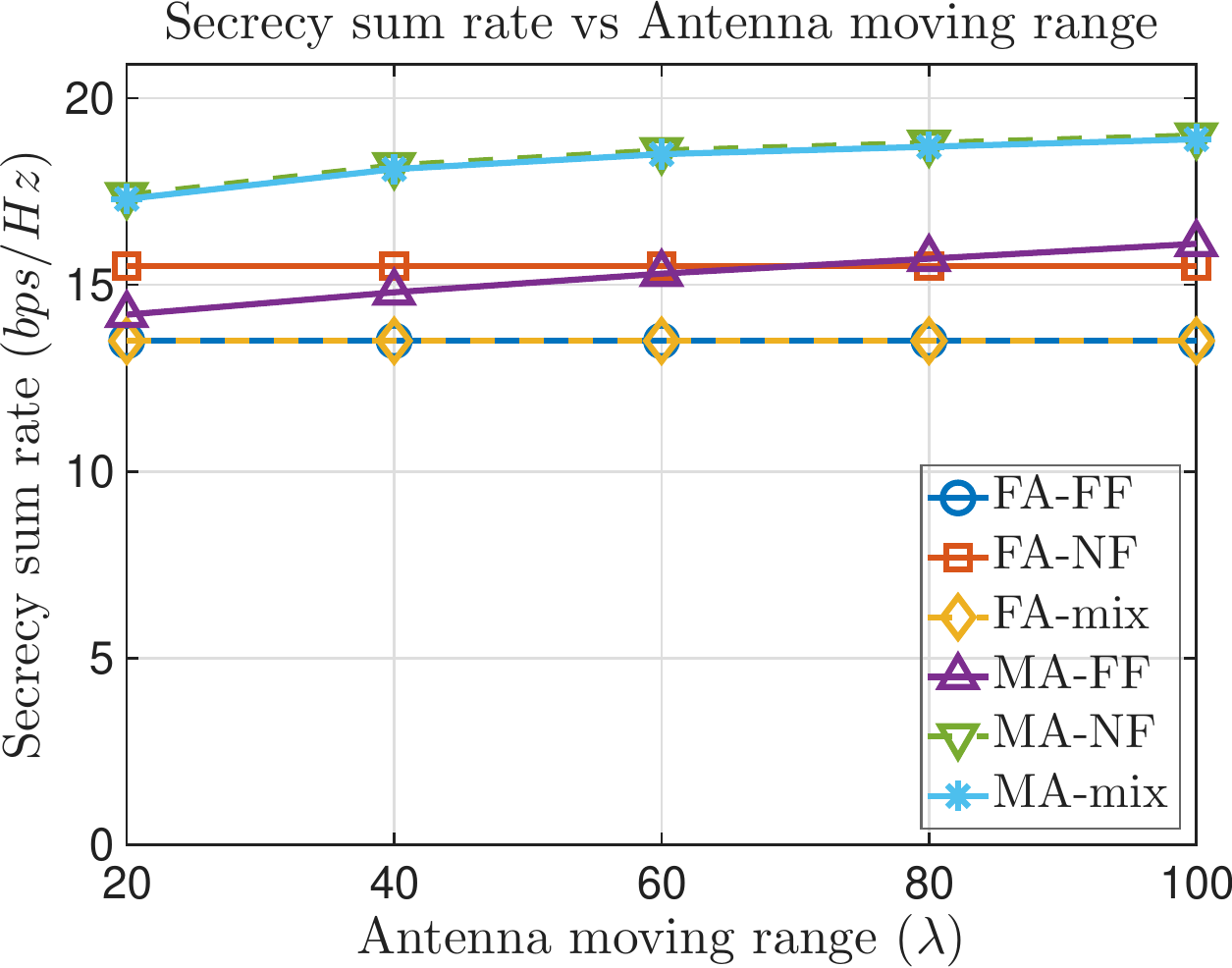}
        \caption{Sum secrecy rate versus antenna moving range.}
        \label{fig:aperture}
    \end{minipage}\hfill
    \begin{minipage}[t]{0.3\textwidth}
        \centering
        \includegraphics[width=\linewidth]{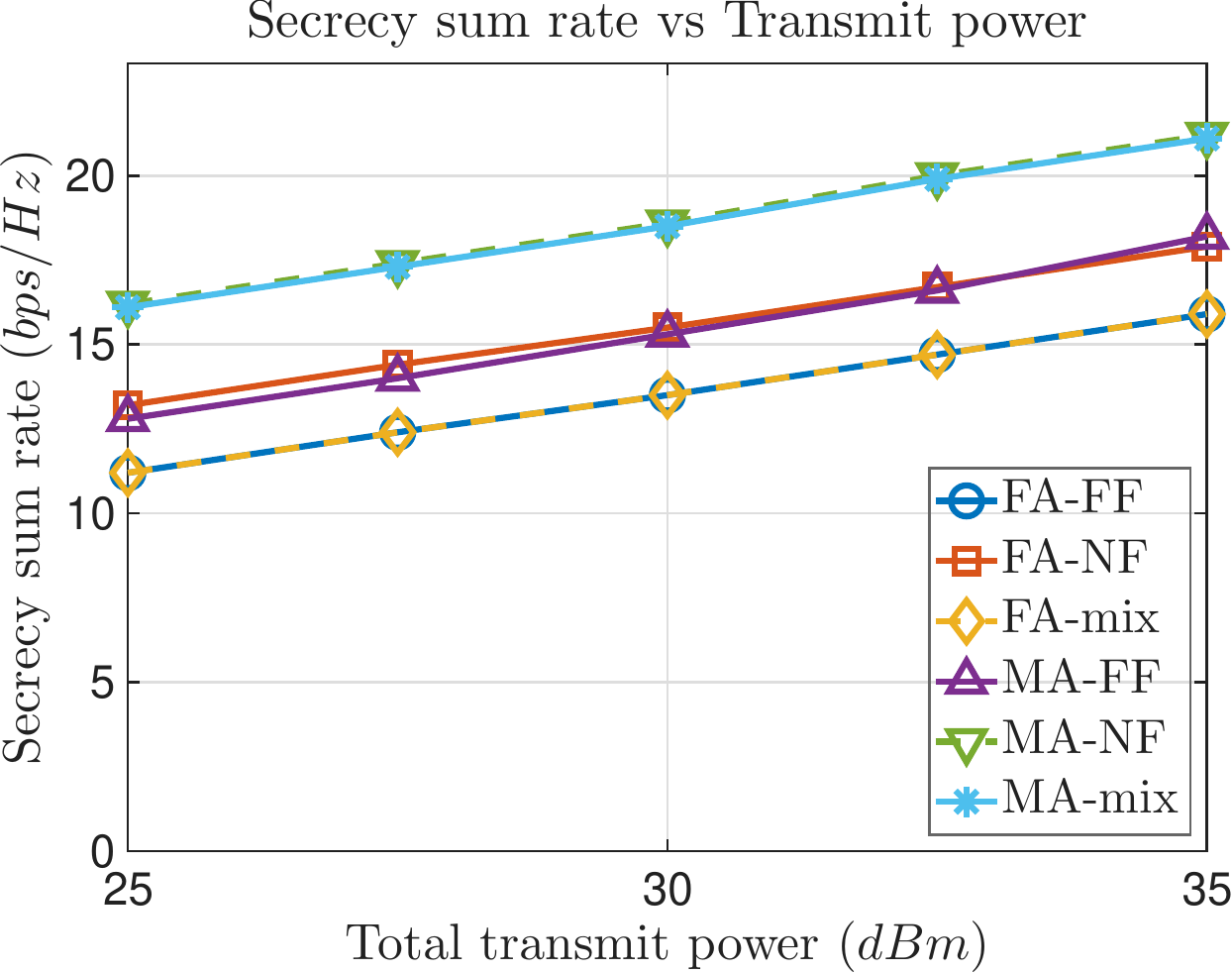}
        \caption{Sum secrecy rate versus total transmit power.}
        \label{fig:power}
    \end{minipage}\hfill
    \begin{minipage}[t]{0.3\textwidth}
        \centering
        \includegraphics[width=\linewidth]{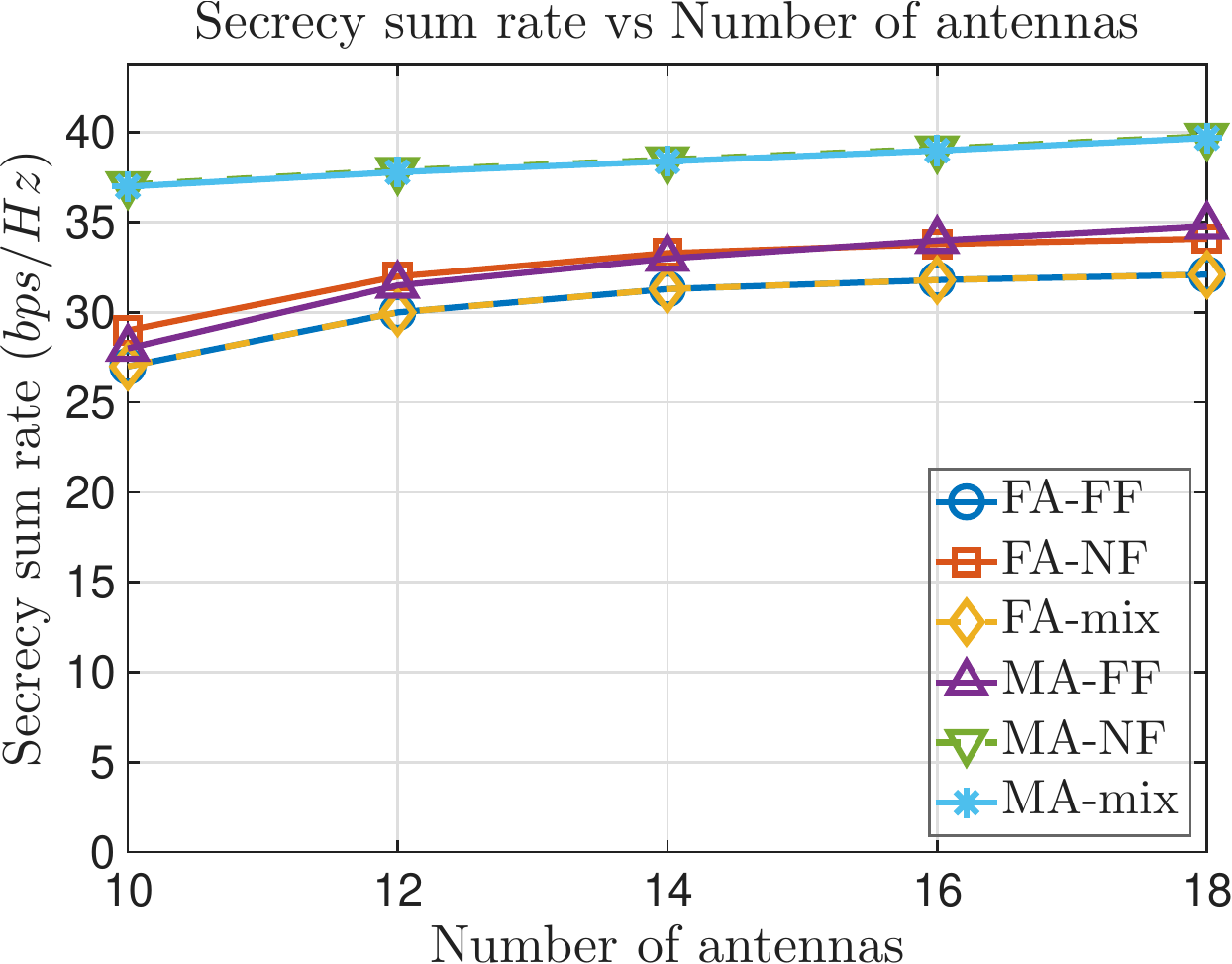}
        \caption{Sum secrecy rate versus number of antennas.}
        \label{fig:antenna}
    \end{minipage}
\end{figure*}

In Figs. \ref{fig:aperture}-\ref{fig:antenna}, two CUs are located at $CU_1(6m,85^{\circ })$, $CU_2(8m,75^{\circ })$. The Eve is located at $E(8m,86^{\circ })$. Two targets are located at $S_1(6m,95^{\circ })$, $S_2(8m,105^{\circ })$, and two clutters are located at $C_1(8m,94^{\circ })$, $C_2(10m,104^{\circ })$.
In this case, the maximum Rayleigh distance is $d_F=0.2625$ m initially. Thus, $CU_1$, $CU_2$, $S_1$, $S_2$, $C_1$, and $C_2$ are in the FF region initially. They stay the same for the FA-based schemes but may change for the MA-based schemes due to antenna position movement.

In Fig. \ref{fig:aperture}, the SSR for different values of the maximum antenna moving range $A=[L_{min},L_{max}]$ is presented. For MA-based schemes, the SSR increases with $A$ and saturates beyond $60\lambda$, whereas FA-based schemes remain constant since the antennas are fixed. MA-NF achieves the highest SSR, with MA-mix nearly identical to MA-NF, confirming that the mixed-field model achieves similar performance to NF model but at lower complexity. MA-FF also improves with $A$, consistently outperforming FA-FF and even surpassing FA-NF at large $A$, but it remains below MA-mix and MA-NF. Among FA schemes, FA-FF and FA-mix perform similarly because all devices remain in the FF region with fixed apertures, while FA-NF performs slightly better because the NF model utilizes distance information to distinguish between $CU_1$ and Eve.

In Fig. \ref{fig:power}, the SSR for different values of the transmit power is presented. The SSR increases with transmit power for all schemes. MA-NF and MA-mix are nearly identical and achieve the highest SSR, confirming that the mixed-field model is sufficient. MA-FF remains above FA-FF and reaches FA-NF at high power, but stays below MA-mix and MA-NF since it utilizes the mismatched FF model for devices entering the NF region. Among FA schemes, FA-FF and FA-mix produce the same SSR, while FA-NF is slightly higher for the same range-resolution reason as in Fig. \ref{fig:aperture}.

Fig. \ref{fig:antenna} shows the SSR versus the number of antennas $Q$ for four CUs placed as in Fig. \ref{fig:CU_impact}. The SSR improves with $Q$ for all schemes, since additional antennas provide more spatial DoF for serving CUs and suppressing the Eve. MA-mix and MA-NF are nearly identical and achieve the highest SSR, followed by MA-FF, which outperforms FA-FF for all $Q$ and exceeds FA-NF at large $Q$. FA-FF and FA-mix produce the same SSR, while FA-NF is slightly higher, consistent with Fig. \ref{fig:aperture}.

Overall, the MA-based schemes outperform the FA-based schemes in most cases considered, while using the correct propagation model for antenna position optimization further enhances the MA gain.

\subsection{Effect of Numbers of $M$ and $N$}
\begin{figure*}[!t]
    \centering
    \captionsetup{font={small}}
    \begin{minipage}[t]{0.3\textwidth}
        \centering
        \includegraphics[width=\linewidth]{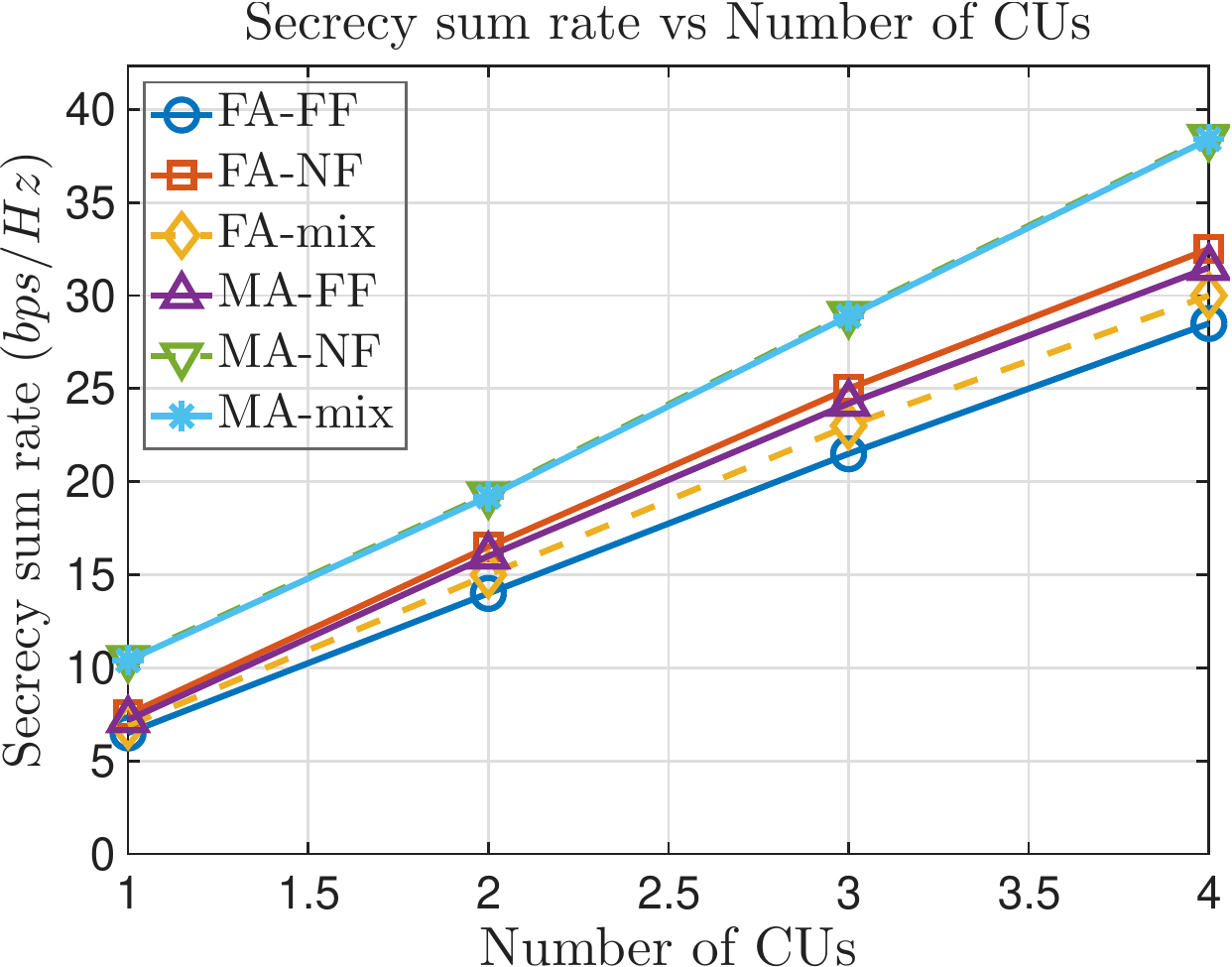}
        \caption{The impact of the number of CUs on sum secrecy rate.}
        \label{fig:CU_impact}
    \end{minipage}\hfill
    \begin{minipage}[t]{0.3\textwidth}
        \centering
        \includegraphics[width=\linewidth]{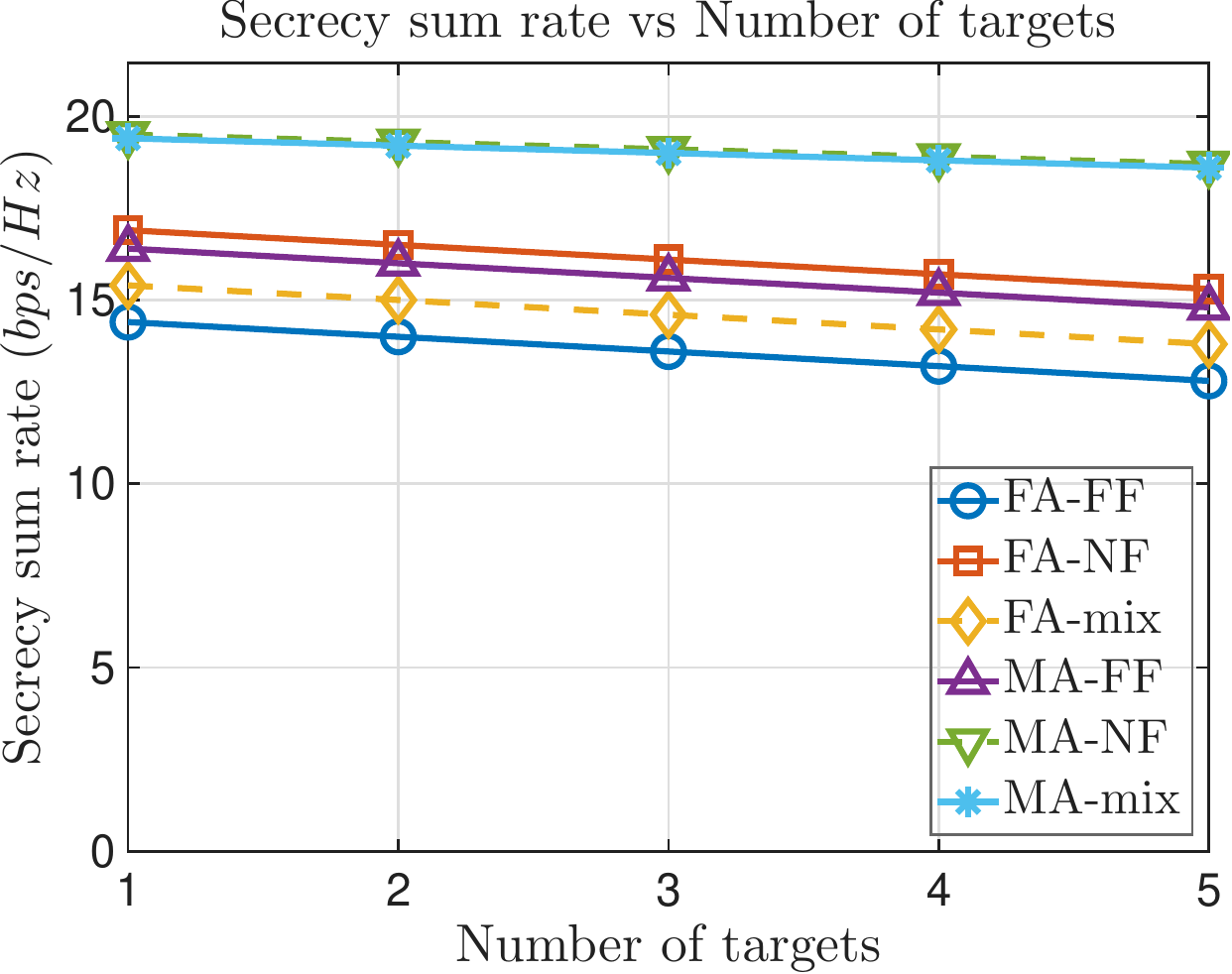}
        \caption{The impact of the number of targets on sum secrecy rate.}
        \label{fig:target_impact}
    \end{minipage}\hfill
    \begin{minipage}[t]{0.3\textwidth}
        \centering
        \includegraphics[width=\linewidth]{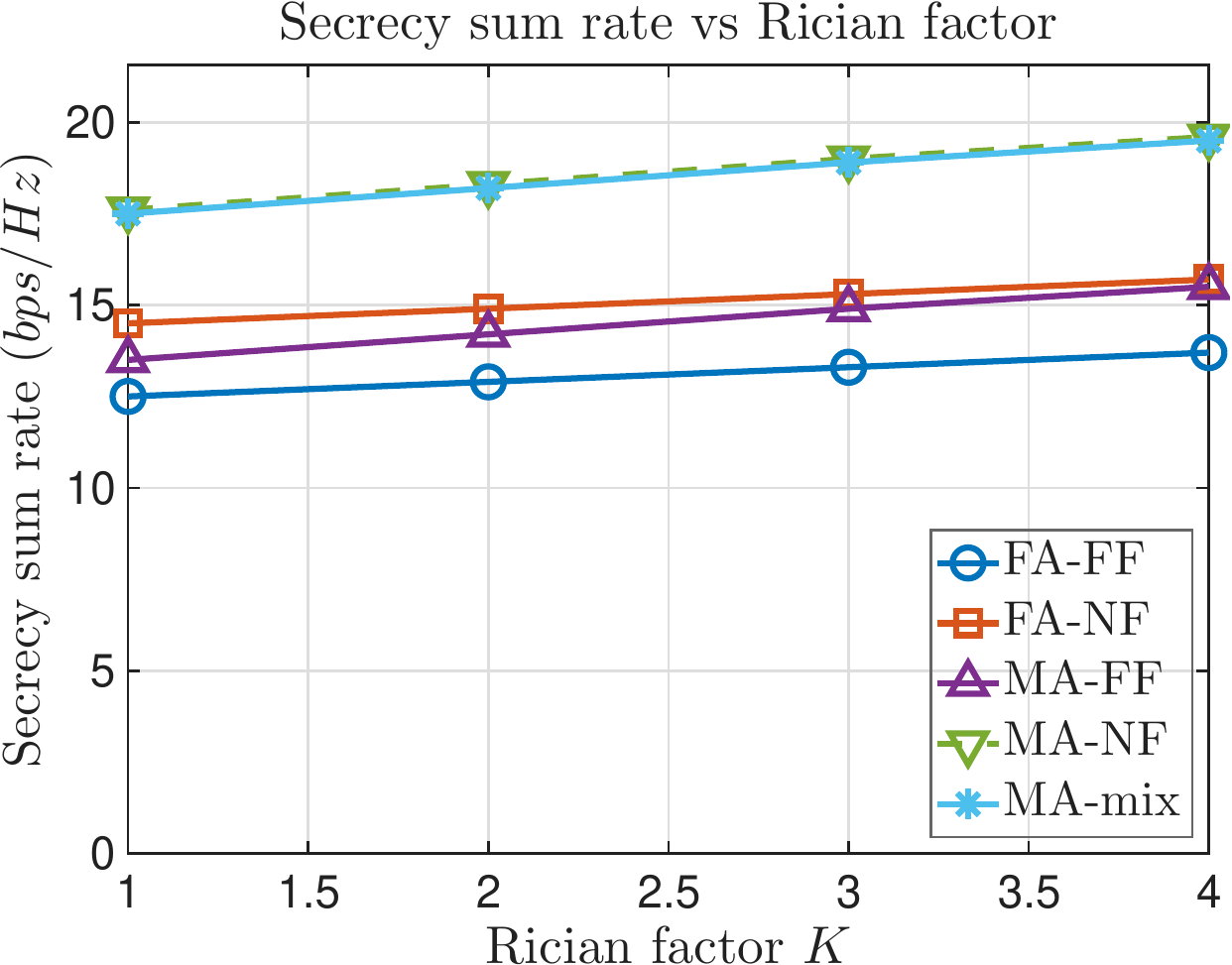}
        \caption{The impact of the Rician factor on sum secrecy rate.}
        \label{fig:Rician factor}
    \end{minipage}
\end{figure*}

Figs. \ref{fig:CU_impact} and \ref{fig:target_impact} reveal the tradeoff between communications and sensing. To show the performance loss caused by the propagation model mismatch, we set $N_t=N_r=40$ antennas. In this case, the Rayleigh distance at the angles of $85^{\circ }$ and $95^{\circ }$ is $6.5375$ meters, indicating that $CU_1$ and $S_1$ are located in the NF while other CUs and targets are in the FF initially. 

Fig. \ref{fig:CU_impact} shows how the number of CUs $M$ affects the SSR. In Fig. \ref{fig:CU_impact}, two additional CUs are considered, located at $CU_3(10m,65^{\circ })$ and $CU_4(12m,55^{\circ })$. For all schemes, SSR increases with the number of CUs, while the average secrecy rate per user decreases due to the power constraint and inter-user interference. MA-NF and MA-mix have the highest SSR and are almost identical. MA-FF has a lower SSR and its performance is comparable to FA-NF. Among the FA schemes, FA-NF has the highest SSR, followed by FA-mix, and finally FA-FF. This is because $CU_1$ and $S_1$ are located in the NF region, and applying the NF model to them can utilize the distance information.

Fig. \ref{fig:target_impact} shows how the number of targets $N$ affects the SSR. In this figure, $CU_1$ and $CU_2$ are considered. Three additional targets are included, located at $S_3(10m,115^{\circ })$, $S_4(12m,125^{\circ })$, and $S_5(14m,135^{\circ })$, while three additional clutters are located at $C_3(12m,114^{\circ })$, $C_4(14m,124^{\circ })$, and $C_5(16m,134^{\circ })$. 
The SSR decreases with the number of targets for all schemes, but the MA-based schemes decrease more slowly because of the antenna position optimization. MA-NF and MA-mix achieve the highest SSR with nearly identical performance. MA-FF is slightly below MA-NF but close to FA-NF, and both outperform FA-mix and FA-FF. FA-mix consistently achieves intermediate performance between FA-FF and FA-NF.

\subsection{Effect of the $K$ and $L$}
Fig. \ref{fig:Rician factor} illustrates the effect of the Rician factor $\mathcal{K}$ on the SSR, where $L=3$. Since FA-mix overlaps with FA-FF when all devices are in the FF region, FA-mix is omitted. For all schemes, SSR increases with $\mathcal{K}$ because a stronger LoS component causes the beamforming gain to concentrate more towards the CU direction. MA-NF and MA-mix have the highest and almost identical SSR, while MA-FF is consistently higher than FA-FF and approaches FA-NF with increasing $\mathcal{K}$. The reason FA-NF is higher than FA-FF is that when $CU_1$ and Eve have similar angles but different distances, the NF model uses range domain information to distinguish them.

\begin{figure}
    \centering
    \includegraphics[width=0.32\textwidth]{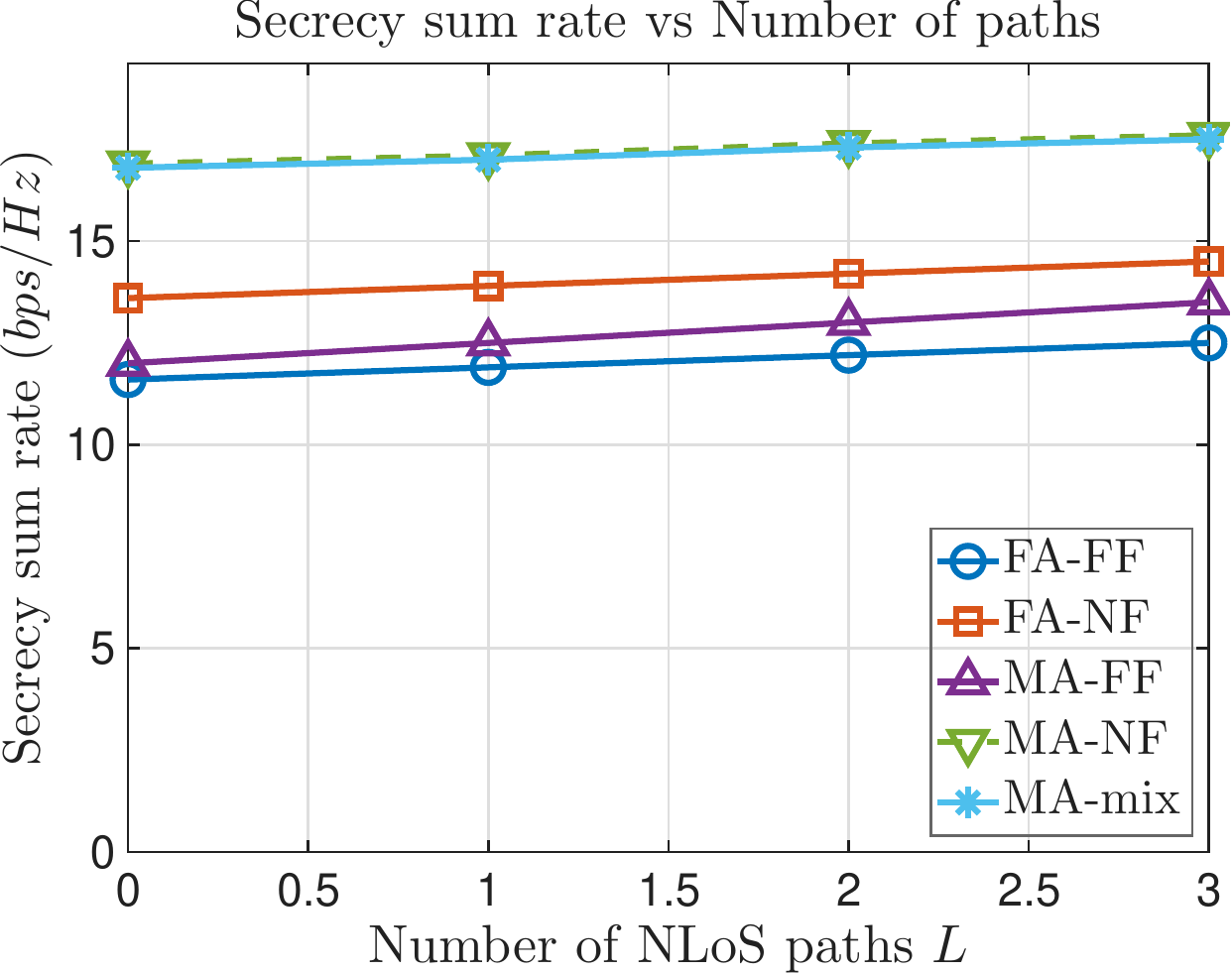}
    \captionsetup{font={small}} 
    \caption{The impact of the number of paths on sum secrecy rate.}
    \label{fig:multipaths}
\end{figure}

Fig. \ref{fig:multipaths} illustrates the effect of the number of paths $L$ on SSR, where $\mathcal{K}=1$. For all schemes, SSR increases with $L$, because richer scattering reduces the correlation between the CU and Eve channels. MA-NF and MA-mix consistently have the highest and almost identical SSR. MA-FF has a higher SSR than FA-FF but lower than FA-NF, and its SSR increases the fastest with increasing $L$, as the additional paths partially compensate for the FF model's difficulty in separating CUs with similar angles.

\subsection{Convergence performance}
\begin{figure*}[!t] 
	\centering
	\begin{subfigure}[t]{0.25\textwidth}
		\centering
		\includegraphics[width=\linewidth]{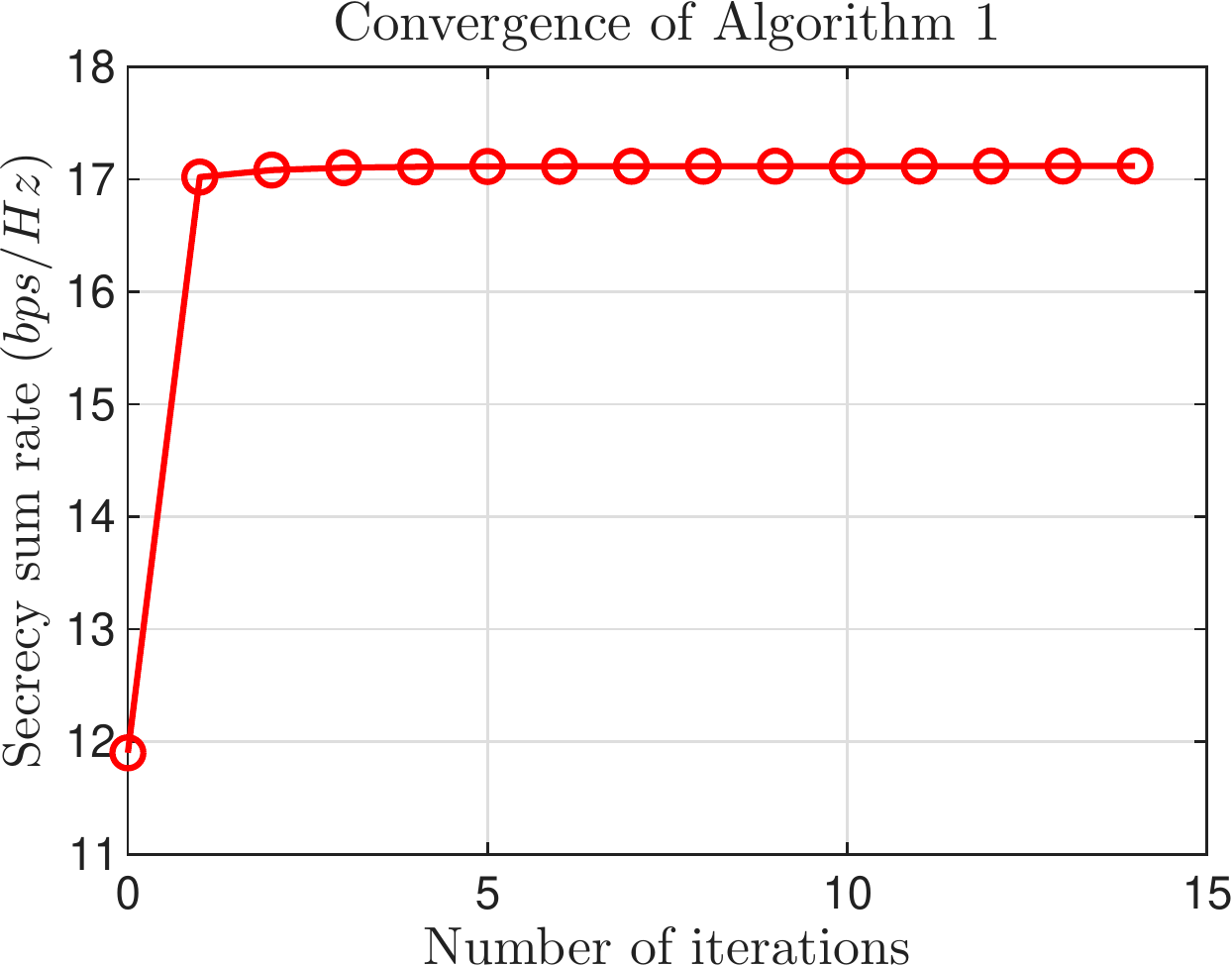}
		\caption{Convergence performance of the Algorithm 1.}
		\label{fig:convergence(a)}
	\end{subfigure}\hfill
	\begin{subfigure}[t]{0.26\textwidth}
		\centering
		\includegraphics[width=\linewidth]{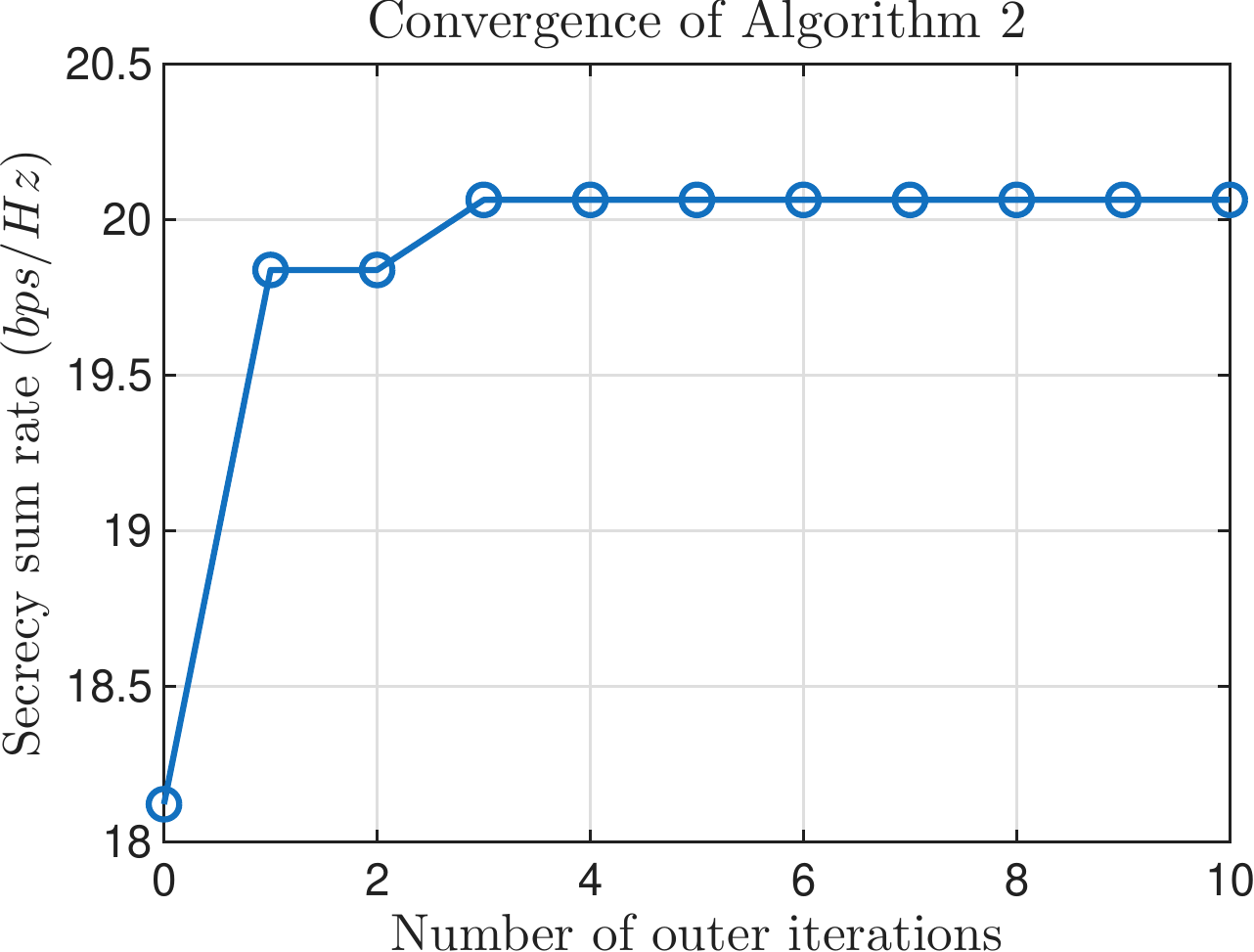}
		\caption{Convergence performance of the Algorithm 2.}
		\label{fig:convergence(b)}
	\end{subfigure}\hfill
	\begin{subfigure}[t]{0.25\textwidth}
		\centering
		\includegraphics[width=\linewidth]{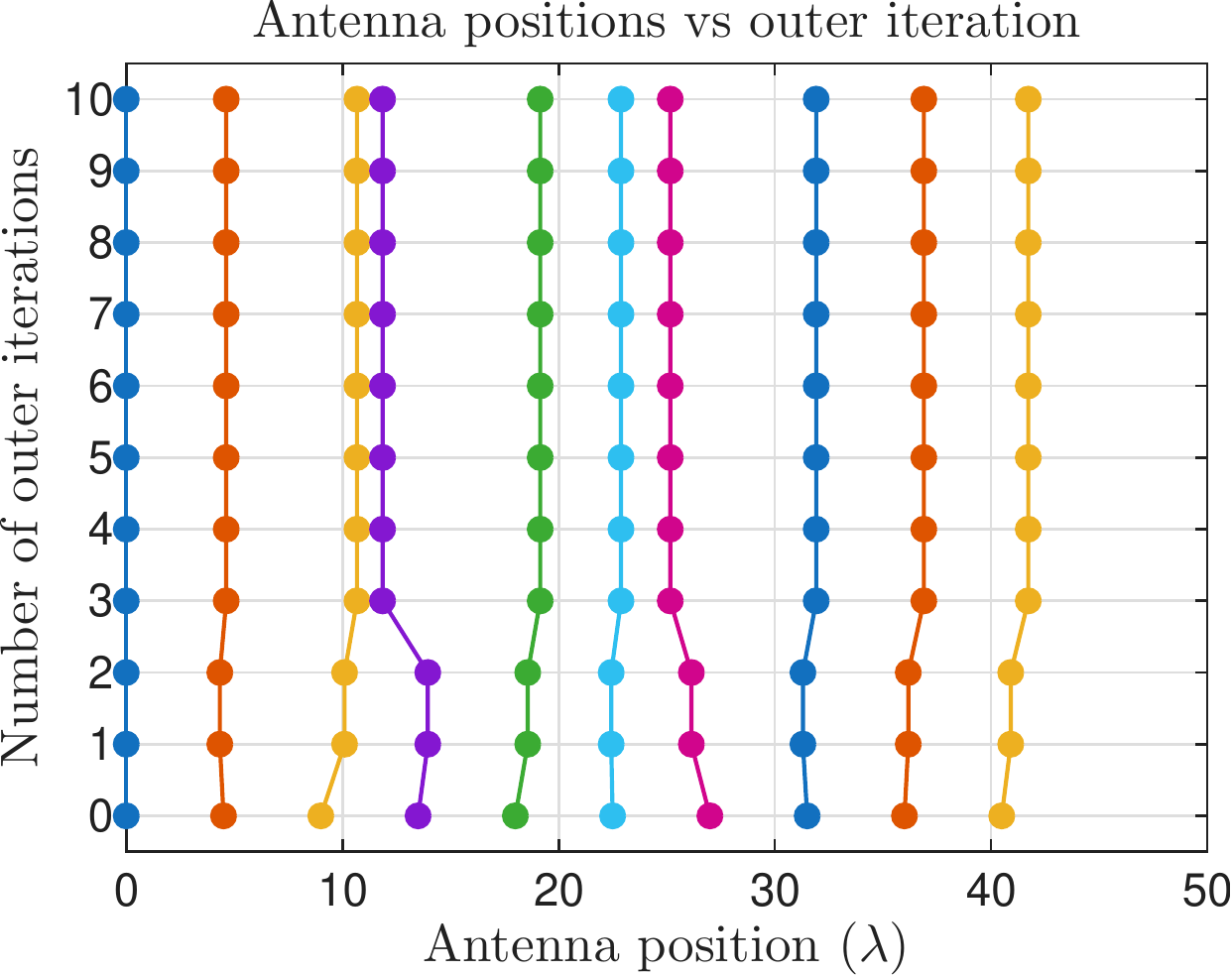}
		\caption{Antenna positions versus the number of outer iterations.}
		\label{fig:convergence(c)}
	\end{subfigure}
	\caption{Convergence performances of proposed algorithms and the updated antenna positions.}
	\label{fig:convergence}
\end{figure*}

This section presents the convergence performance of the proposed algorithms. To avoid randomness introduced by the randomly generated multipath components, only the LoS channels are considered. 
Fig. \ref{fig:convergence}(a) and Fig. \ref{fig:convergence}(b) show the convergence performance of Algorithm \ref{algorithm 1} and Algorithm \ref{algorithm 2} in a mixed-field environment, respectively, with $N_t=N_r=10$ antennas, $P_0=30$ dBm, and $A=60\lambda$. $CU_1$, $CU_2$, $S_1$, $S_2$, $C_1$, and $C_2$ are considered.
Fig. \ref{fig:convergence}(a) shows the convergence performance of the beamforming design algorithm in Algorithm \ref{algorithm 1}. It can be observed that the SSR increases monotonically with the number of iterations, which demonstrates the effectiveness of the proposed beamforming design. Fig. \ref{fig:convergence}(b) shows the convergence performance of Algorithm \ref{algorithm 2} with antenna position update. In the first two outer iterations, the SSR gradually increases and then stabilizes at approximately 20.06 bps/Hz. Due to the line search and precise projection in Steps 4) and 5), the SSR is non-decreasing. The BB step adjusts the step size according to the local curvature, avoiding the oscillations around the optimum caused by a fixed step size with the sequential clamping projection. Fig. \ref{fig:convergence}(c) shows the antenna positions after each outer iteration. Starting with an initial uniform array, the positions are adjusted in the first two iterations and then tend to stabilize, which is consistent with the SSR curve in Fig. \ref{fig:convergence}(b).

\begin{figure}[!t]
    \centering
    \includegraphics[width=0.25\textwidth]{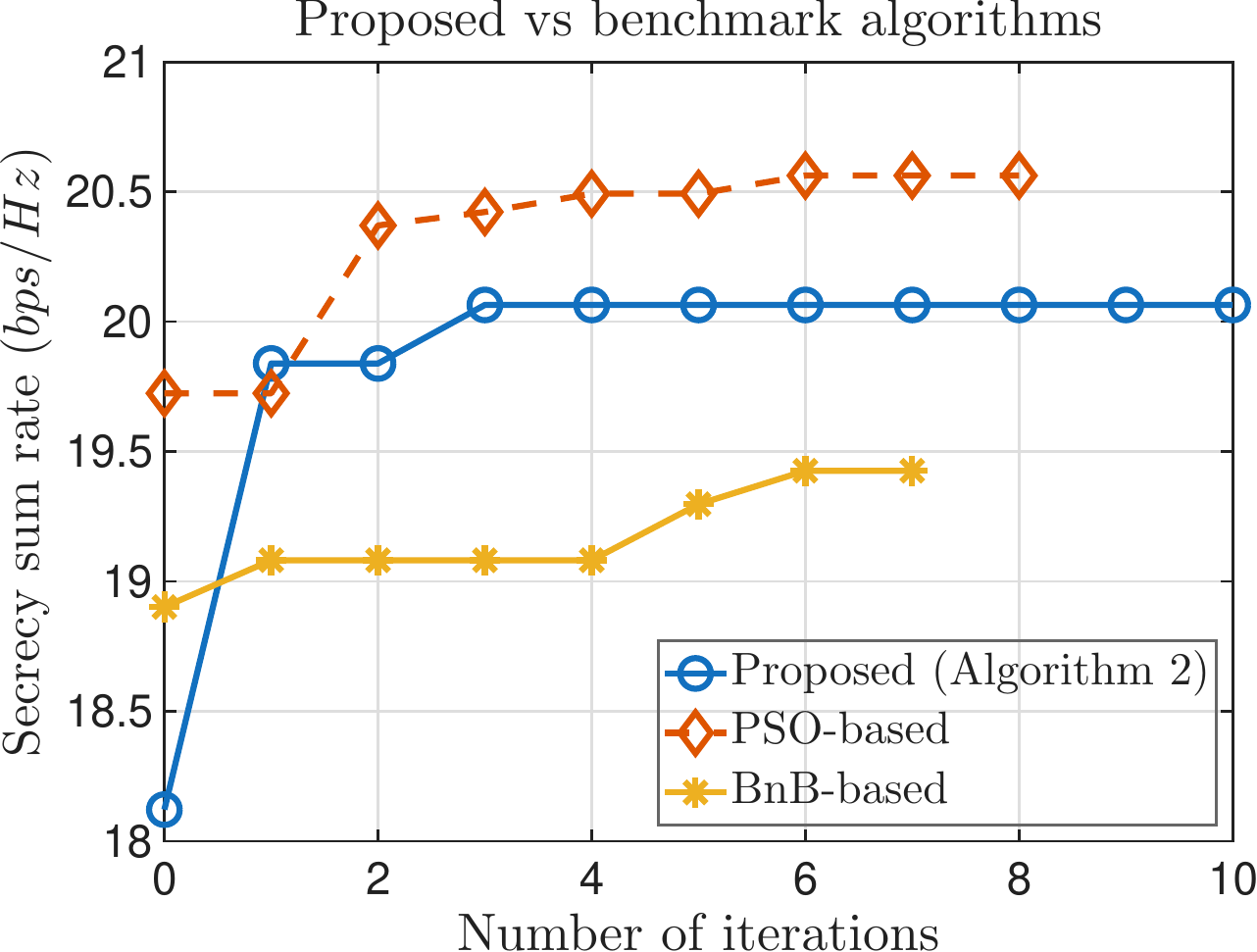}
    \captionsetup{font={small}}
    \caption{SSR versus the number of iterations for Algorithm~2 and the PSO- and BnB-based benchmarks, with $N_t=N_r=10$, $A=60\lambda$, and LoS channels.}
    \label{fig:baseline}
\end{figure}

Fig.~\ref{fig:baseline} and Table~\ref{tab:runtime} compare the proposed Algorithm~\ref{algorithm 2} with the particle swarm optimization (PSO)- and BnB-based benchmarks under the same setup, with all methods using Algorithm~\ref{algorithm 1} for beamforming. The PSO benchmark uses particle swarm optimization to search for antenna positions, where each particle is a candidate position vector with fitness derived from the optimized SSR \cite{kennedy1995particle}. The BnB benchmark, on the other hand, follows discrete antenna localization of \cite{11048972} and searches for candidate uniform spacings in (\ref{eq:spacing set}) using a branch-and-bound method. As shown in Fig.~\ref{fig:baseline}, all three methods converge within a few iterations to comparable SSRs. The total complexity is determined mainly by: one per outer iteration for the proposed algorithm, whose line search uses the closed-form $\Lambda(\mathbf{x})$, $N_{\mathrm{p}}T_{\mathrm{p}}$ for PSO, and $N_{\mathrm{ev}}$ for BnB. The measured runtimes are consistent with this analysis. PSO achieves a slightly higher SSR than the proposed method ($20.56$ vs. $20.06$~bps/Hz, a $2.5\%$ gain) but requires $291.1$~s compared with $166.6$~s. BnB is faster ($48.7$~s) but yields the lowest SSR ($19.43$~bps/Hz) due to its restriction to uniform spacings. Overall, the proposed method provides the best quality--complexity trade-off.

\section{Conclusion}\label{Sec:conclusion}
This paper has investigated the PLS performance of the MA-aided ISAC in a mixed near- and far-field environment. A joint optimization problem has been formulated to maximize the SSR by designing transmit beamforming and antenna positions, while accounting for propagation model updates caused by changes in the Rayleigh distance during antenna position adjustments. 
Numerical results have validated that the SSR increases with the moving range of antennas, the transmit power, the number of antennas, the Rician factor, and the number of multipath components, while it decreases with the number of sensing targets.
They have also addressed the challenges raised in the Introduction. Specifically, MA-NF that always uses the NF models outperforms MA-mix that uses the actual models during antenna position update, but the performance difference is marginal. Hence, it is not beneficial to always use NF models in a mixed-field scenario, as the channel estimation for NF models is much more complicated than that for FF models, especially with large arrays. 
Furthermore, ignoring the propagation model changes caused by antenna position updates results in a loss of MA gain and, in some cases, performance that is inferior to FA-based ISAC using the correct model.
Therefore, MA-aided ISAC can obtain its full PLS gain when the propagation field changes are considered during antenna updates.

\appendix[Derivation of the Antenna-Position Gradient]\label{app:gradient}
This appendix presents the full derivation of the closed-form elementary gradients used to assemble $\bigtriangledown_{\mathbf{x}}\Lambda(\mathbf{x})$ in (\ref{eq:Final gradient}).

For the $j$-th CU, define two functions as $\mathcal{G}_{j,k}\in\{\mathcal{G}_{j,k}^{NF},\mathcal{G}_{j,k}^{FF} \}$, $\mathcal{G}_{e,p}\in\{\mathcal{G}_{e,p}^{NF},\mathcal{G}_{e,p}^{FF} \}$, where
\begin{equation}
    \mathcal{G}_{j,k}^{NF}=\textbf{b}_{k}^{H} \mathcal{H}_{j}^{NF}(\textbf{x})\textbf{b}_{k}= \left | (\boldsymbol{\beta}_{j}^{NF}\odot\textbf{a}(\textbf{x},r_j, \theta_{j}))^{H}\textbf{b}_{k} \right |^2,
\end{equation}
\begin{equation}
    \mathcal{G}_{j,k}^{FF}=\textbf{b}_{k}^{H} \mathcal{H}_{j}^{FF}(\textbf{x})\textbf{b}_{k}= \left | (\boldsymbol{\beta}_{j}^{FF}\odot\textbf{a}(\textbf{x}, \theta_{j}))^{H}\textbf{b}_{k} \right |^2
\end{equation}
\begin{equation}
\mathcal{G}_{e,p}^{\mathrm{NF}}
= \mathbf{b}_{p}^{H} \mathcal{H}_{e}^{\mathrm{NF}}(\mathbf{x}) \mathbf{b}_{p}
= \left| (\boldsymbol{\beta}_e^{NF} \odot \mathbf{a}(\mathbf{x}, r_e, \theta_e))^{H} \mathbf{b}_p \right|^2
\end{equation}
\begin{equation}
\mathcal{G}_{e,p}^{\mathrm{FF}}
= \mathbf{b}_{p}^{H} \mathcal{H}_{e}^{\mathrm{FF}}(\mathbf{x}) \mathbf{b}_{p}
= \left| (\boldsymbol{\beta}_e^{\mathrm{FF}} \odot \mathbf{a}(\mathbf{x}, \theta_e))^{H} \mathbf{b}_p \right|^2.
\end{equation}
Then, we define two functions as $\mathcal{F}_{j}\in\{\mathcal{F}_{j}^{NF},\mathcal{F}_{j}^{FF}\}$, and $\mathcal{F}_{ej}\in\{\mathcal{F}_{ej}^{NF},\mathcal{F}_{ej}^{FF}\}$, where $\mathcal{F}_{j}^{NF}=\sqrt{\mathcal{G}_{j,k}^{NF}}$, $\mathcal{F}_{j}^{FF}=\sqrt{\mathcal{G}_{j,k}^{FF}}$, and
\begin{equation}
\mathcal{F}_{ej}^{NF}=\sqrt{\textbf{b}_{j}^{H}\mathcal{H}_{e}^{NF} (\textbf{x})\textbf{b}_{j}}
=\left | (\boldsymbol{\beta}_{e}^{NF}\odot\textbf{a}(\textbf{x},r_e, \theta_{e}))^{H}\textbf{b}_{j} \right |,
\end{equation}
\begin{equation}    \mathcal{F}_{ej}^{FF}=\sqrt{\textbf{b}_{j}^{H}\mathcal{H}_{e}^{FF}} (\textbf{x})\textbf{b}_{j}
=\left | (\boldsymbol{\beta}_{e}^{FF}\odot\textbf{a}(\textbf{x}, \theta_{e}))^{H}\textbf{b}_{j} \right |.
\end{equation}
Thus, the gradient of $\Lambda \left (\textbf{x}  \right ) $ with respect to $x_{n}$ can be expressed as
\begin{equation}\label{eq:gradient1}
\begin{aligned}
    \frac{\partial \Lambda \left (\textbf{x}  \right ) }{\partial x_{n}}=&\sum_{j=1}^{M}\Bigg( 2\vartheta_{j}\sqrt{(1+\gamma_{j})}\frac{\partial \mathcal{F}_{j}}{\partial  x_{n}} -\sum_{k=1}^{M}\vartheta_{k}^{2}\frac{\partial \mathcal{G}_{j,k}}{\partial  x_{n}}-\\
    &2\vartheta_{e,j}\sqrt{(1-\gamma_{e,j})}\frac{\partial \mathcal{F}_{ej}}{\partial  x_{n}}+  \sum_{p=1,p\ne j}^{M}\vartheta_{e,p}^{2}\frac{\partial \mathcal{G}_{e,p}}{\partial  x_{n}}\Bigg ) .
\end{aligned}
\end{equation}
Define $\textbf{g}_{j}\in\left \{ \textbf{g}_{NF,j}\text{, }\textbf{g}_{FF,j} \right \}$, $\textbf{w}_{j}\in\left \{ \textbf{w}_{NF,j}\text{, }\textbf{w}_{FF,j} \right \}$, $\textbf{g}_{e}\in\left \{ \textbf{g}_{NF,e}\text{, }\textbf{g}_{FF,e} \right \}$, and $\textbf{w}_{e}\in\left \{ \textbf{w}_{NF,e}\text{, }\textbf{w}_{FF,e} \right \}$, where $\textbf{g}_{NF,j} = \left [ g_{1,j}^{NF},g_{2,j}^{NF},\dots, g_{N_{t},j}^{NF}\right ]^{T}$, $\textbf{g}_{FF,j} = \left [ g_{1,j}^{FF},g_{2,j}^{FF},\dots, g_{N_{t},j}^{FF}\right ]^{T}$, $\textbf{w}_{NF,j} = \left [ w_{1,j}^{NF},w_{2,j}^{NF},\dots, w_{N_{t},j}^{NF}\right ]^{T}$, $\textbf{w}_{FF,j} = \left [ w_{1,j}^{FF},w_{2,j}^{FF},\dots, w_{N_{t},j}^{FF}\right ]^{T}$, $\textbf{g}_{NF,e} = \left [ g_{1,e}^{NF},g_{2,e}^{NF},\dots, g_{N_{t},e}^{NF}\right ]^{T}$, $\textbf{g}_{FF,e} = \left [ g_{1,e}^{FF},g_{2,e}^{FF},\dots, g_{N_{t},e}^{FF}\right ]^{T}$, $\textbf{w}_{NF,e} = \left [ w_{1,e}^{NF},w_{2,e}^{NF},\dots, w_{N_{t},e}^{NF}\right ]^{T}$, and $\textbf{w}_{FF,e} = \left [ w_{1,e}^{FF},w_{2,e}^{FF},\dots, w_{N_{t},e}^{FF}\right ]^{T}$. The NF parameters $g_{n,j}^{NF}$, $g_{n,e}^{NF}$, $w_{n,j}^{NF}$, and $w_{n,e}^{NF}$ are defined as
\begin{equation}
    g_{n,j}^{NF}=\cos\left ( \frac{2\pi }{\lambda}\left ( r_{(x_{n})}-r_{j} \right )  \right ),
    w_{n,j}^{NF}=\sin\left ( \frac{2\pi }{\lambda}\left ( r_{(x_{n})}-r_{j} \right )  \right )
\end{equation}
\begin{equation}
    g_{n,e}^{NF}=\cos\left ( \frac{2\pi }{\lambda}\left ( r_{(x_{n})}-r_{e} \right )  \right ),
    w_{n,e}^{NF}=\sin\left ( \frac{2\pi }{\lambda}\left ( r_{(x_{n})}-r_{e} \right )  \right ).
\end{equation}
The FF parameters $g_{n,j}^{FF}$, $g_{n,e}^{FF}$, $w_{n,j}^{FF}$, and $w_{n,e}^{FF}$ are defined as
\begin{equation}
    g_{n,j}^{FF}= \cos \left ( \frac{2\pi}{\lambda}x_{n}\cos (\theta_{j})  \right ), w_{n,j}^{FF}= \sin \left ( \frac{2\pi}{\lambda}x_{n}\cos (\theta_{j})  \right ),
\end{equation}
\begin{equation}
    g_{n,e}^{FF}= \cos \left ( \frac{2\pi}{\lambda}x_{n}\cos (\theta_{e})  \right ), w_{n,e}^{FF}= \sin \left ( \frac{2\pi}{\lambda}x_{n}\cos (\theta_{e})  \right ).
\end{equation}

By applying the Euler formula to the beam focusing/steering vectors, we can have $\textbf{a}(\mathbf{x},\theta,\textbf{r})= \textbf{g}_{NF}-J\textbf{w}_{NF}$ and $\textbf{a}(\mathbf{x},\theta)=\textbf{g}_{FF}+J\textbf{w}_{FF}$, where $J$ denotes the imaginary unit.
Denote $\textbf{b}_{k}=\textbf{u}_{k}+J\textbf{v}_{k}$, $\Theta_{k}=\textbf{u}_{k}\textbf{u}_{k}^{T}+\textbf{v}_{k}\textbf{v}_{k}^{T}$, $\Phi_{k}=\textbf{u}_{k}\textbf{v}_{k}^{T}-\textbf{v}_{k}\textbf{u}_{k}^{T}$. For CUs, we can have
\begin{equation}\label{f(g,w) NF}
\begin{aligned}
\mathcal{G}_{j,k}^{NF}
&=\textbf{g}_{NF,j}^{T}\left ( \Upsilon_j^{NF}\odot\Theta_{k} \right ) \textbf{g}_{NF,j}
+\textbf{w}_{NF,j}^{T}\left ( \Upsilon_j^{NF}\odot\Theta_{k} \right )\textbf{w}_{NF,j} \\
&\quad -2\textbf{g}_{NF,j}^{T}\left ( \Upsilon_j^{NF}\odot\Phi_{k} \right )\textbf{w}_{NF,j}
= f_{NF}\left ( \textbf{g}_{NF,j},\textbf{w}_{NF,j} \right ),
\end{aligned}
\end{equation}
\begin{equation}\label{f(g,w) FF}
\begin{aligned}
   \mathcal{G}_{j,k}^{FF}&=\textbf{g}_{FF,j}^{T}\left ( \Upsilon_j^{FF}\odot\Theta_{k} \right )  \textbf{g}_{FF,j}+\textbf{w}_{FF,j}^{T}\left( \Upsilon_j^{FF}\odot\Theta_{k} \right )\textbf{w}_{FF,j} \\
    &+2\textbf{g}_{FF,j}^{T}\left ( \Upsilon_j^{FF}\odot\Phi_{k}\right)\textbf{w}_{FF,j}=f_{FF}\left ( \textbf{g}_{FF,j},\textbf{w}_{FF,j} \right ),
\end{aligned}
\end{equation}
where $\Upsilon_j^{NF} =(\boldsymbol{\beta}_j^{\mathrm{NF}} (\boldsymbol{\beta}_j^{\mathrm{NF}})^H)$ and $\Upsilon_j^{FF} =(\boldsymbol{\beta}_j^{\mathrm{FF}} (\boldsymbol{\beta}_j^{\mathrm{FF}})^H)$. Then, the gradient of $\mathcal{G}_{j,k}^{NF}$ and $\mathcal{G}_{j,k}^{FF}$ with respect to the $n$-th antenna position can be calculated as (\ref{Deri_f_NF}) and (\ref{Deri_f_FF}) on the top of this page.
\begin{figure*} 
    \centering
    \vspace*{-\baselineskip}
\begin{equation}\label{Deri_f_NF}
    \begin{aligned}
        &\frac{\partial f_{NF}\left ( \textbf{g}_{NF,j},\textbf{w}_{NF,j} \right ) }{\partial x_n}=\frac{\partial f\left ( \textbf{g}_{NF,j},\textbf{w}_{NF,j} \right ) }{\partial g_{n,j}^{NF}}\frac{\partial g_{n,j}^{NF} }{x_n}+
        \frac{\partial f\left ( \textbf{g}_{NF,j},\textbf{w}_{NF,j} \right ) }{\partial w_{n,j}^{NF}}\frac{\partial w_{n,j}^{NF} }{x_n}=\\
        &-\frac{\partial f\left ( \textbf{g}_{NF,j},\textbf{w}_{NF,j} \right ) }{\partial g_{n,j}^{NF}} \frac{2\pi }{\lambda}\frac{x_{n}-r_{j}\cos(\theta_{j})}{r_{(x_{n})}} \sin\left ( \frac{2\pi }{\lambda}\left ( r_{(x_{n})}-r_{j} \right )  \right ) +
         \frac{\partial f\left ( \textbf{g}_{NF,j},\textbf{w}_{NF,j} \right ) }{\partial w_{n,j}^{NF}} \frac{2\pi }{\lambda}\frac{x_{n}-r_{j}\cos(\theta_{j})}{r_{(x_{n})}} \cos\left ( \frac{2\pi }{\lambda}\left ( r_{(x_{n})}-r_{j} \right )  \right )
    \end{aligned}
\end{equation}
\hrulefill 
\end{figure*}
\begin{figure*}[!t] 
\centering 
\vspace*{-\baselineskip}
\begin{equation}\label{Deri_f_FF}
    \begin{aligned}
    &\frac{\partial f_{FF}\!\left( \textbf{g}_{FF,j}, \textbf{w}_{FF,j} \right)}{\partial x_{n}}
    = \frac{\partial f_{FF}\!\left( \textbf{g}_{FF,j}, \textbf{w}_{FF,j} \right)}{\partial g_{n,j}^{FF}}
        \frac{\partial g_{n,j}^{FF}}{\partial x_{n}}
     +  \frac{\partial f_{FF}\!\left( \textbf{g}_{FF,j}, \textbf{w}_{FF,j} \right)}{\partial w_{n,j}^{FF}}
        \frac{\partial w_{n,j}^{FF}}{\partial x_{n}} \\[4pt]
    &= -\,\frac{\partial f_{FF}\!\left( \textbf{g}_{FF,j}, \textbf{w}_{FF,j} \right)}{\partial g_{n,j}^{FF}}
         \frac{2\pi}{\lambda}\cos(\theta_{j})
         \sin\!\left( \frac{2\pi}{\lambda}x_{n}\cos(\theta_{j}) \right)  +\,\frac{\partial f_{FF}\!\left( \textbf{g}_{FF,j}, \textbf{w}_{FF,j} \right)}{\partial w_{n,j}^{FF}}
         \frac{2\pi}{\lambda}\cos(\theta_{j})
         \cos\!\left( \frac{2\pi}{\lambda}x_{n}\cos(\theta_{j}) \right) 
    \end{aligned}
\end{equation}
\hrulefill 
\end{figure*}
Given $f\left ( \textbf{g}_{j},\textbf{w}_{j} \right )\in\left \{ f_{NF}\left ( \textbf{g}_{NF,j},\textbf{w}_{NF,j} \right ),f_{FF}\left ( \textbf{g}_{FF,j},\textbf{w}_{FF,j} \right ) \right \}    $ in (\ref{f(g,w) NF}) and (\ref{f(g,w) FF}), we can obtain 
\begin{equation}
\begin{aligned}
    &\bigtriangledown_{\textbf{g}_{FF,j}} f_{FF}\left ( \textbf{g}_{FF,j},\textbf{w}_{FF,j} \right )
=\\
&2\left(\Upsilon_j^{FF}\odot\Theta_{k}\right)\textbf{g}_{FF,j}
+2\left(\Upsilon_j^{FF}\odot\Phi_{k}\right)\textbf{w}_{FF,j},
\end{aligned}
\end{equation}
\begin{equation}
\begin{aligned}
&\bigtriangledown_{\textbf{w}_{FF,j}} f_{FF}\left ( \textbf{g}_{FF,j},\textbf{w}_{FF,j} \right )=\\
&2\left(\Upsilon_j^{FF}\odot\Theta_{k}\right)\textbf{w}_{FF,j}
-2\left(\Upsilon_j^{FF}\odot\Phi_{k}\right)\textbf{g}_{FF,j},
\end{aligned}
\end{equation}

\begin{equation}
\begin{aligned}
&\bigtriangledown_{\textbf{g}_{NF,j}} f_{NF}\left ( \textbf{g}_{NF,j},\textbf{w}_{NF,j} \right )=\\
&
2\left(\Upsilon_j^{NF}\odot\Theta_k\right)\mathbf{g}_{NF,j}
-
2\left(\Upsilon_j^{NF}\odot\Phi_k\right)\mathbf{w}_{NF,j}
,
\end{aligned}
\end{equation}
\begin{equation}
\begin{aligned}
&\bigtriangledown_{\textbf{w}_{NF,j}} f_{NF}\left ( \textbf{g}_{NF,j},\textbf{w}_{NF,j} \right )=\\
&
2\left(\Upsilon_j^{NF}\odot\Theta_k\right)\mathbf{w}_{NF,j}
+
2\left(\Upsilon_j^{NF}\odot\Phi_k\right)\mathbf{g}_{NF,j}.
\end{aligned}
\end{equation}

For FF CUs, denote $\textbf{G}_{FF,j}=\text{diag} \left ( \left \{ \frac{2\pi}{\lambda } \cos(\theta_{j})\sin\left (  \frac{2\pi}{\lambda } x_{n}\cos(\theta_{j})\right ) \right \}_{n=1}^{N_{t}}  \right )$ and $\textbf{W}_{FF,j}=\text{diag} \left ( \left \{ \frac{2\pi}{\lambda } \cos(\theta_{j})\cos\left (  \frac{2\pi}{\lambda } x_{n}\cos(\theta_{j})\right ) \right \}_{n=1}^{N_{t}}  \right )$.
Hence, it can be obtained that
\begin{equation}
\begin{aligned}
    &\bigtriangledown_{\textbf{x}}\mathcal{G}_{j,k}^{FF}=\bigtriangledown_{\textbf{x}}f_{FF}\left ( \textbf{g}_{FF,j},\textbf{w}_{FF,j} \right )=\\
&-\textbf{G}_{j}^{FF}\Big ( 2(\Upsilon_j^{FF}\odot\Theta_{k}) \textbf{g}_{FF,j}
+2(\Upsilon_j^{FF}\odot\Phi_{k})\textbf{w}_{FF,j} \Big )
\\
&+\textbf{W}_{j}^{FF}\Big ( 2(\Upsilon_j^{FF}\odot\Theta_{k}) \textbf{w}_{FF,j}
-2(\Upsilon_j^{FF}\odot\Phi_{k})\textbf{g}_{FF,j} \Big )
\end{aligned}
\end{equation}

For NF CUs, denote
$ \textbf{G}_{NF,j}=\text{diag} \left ( \left \{ \frac{2\pi }{\lambda}\frac{x_{n}-r_{j}\cos(\theta_{j})}{r_{(x_{n})}} \sin\left ( \frac{2\pi }{\lambda}\left ( r_{(x_{n})}-r_{j} \right )  \right )\right \}_{n=1}^{N_{t}}  \right )$, and $\textbf{W}_{NF,j}=\text{diag} \left ( \left \{ \frac{2\pi }{\lambda}\frac{x_{n}-r_{j}\cos(\theta_{j})}{r_{(x_{n})}} \cos\left ( \frac{2\pi }{\lambda}\left ( r_{(x_{n})}-r_{j} \right )  \right ) \right \}_{n=1}^{N_{t}}  \right )$.
Then, it can be derived that
\begin{equation}
\begin{aligned}
    &\bigtriangledown_{\textbf{x}}\mathcal{G}_{j,k}^{NF}=\bigtriangledown_{\textbf{x}}f_{NF}\left ( \textbf{g}_{NF,j},\textbf{w}_{NF,j} \right )=\\
&-\textbf{G}_{j}^{NF}\Big ( 2(\Upsilon_j^{NF}\odot\Theta_{k}) \textbf{g}_{NF,j}
-2(\Upsilon_j^{NF}\odot\Phi_{k})\textbf{w}_{NF,j} \Big )\\
&
+\textbf{W}_{j}^{NF}\Big ( 2(\Upsilon_j^{NF}\odot\Theta_{k}) \textbf{w}_{NF,j}+2(\Upsilon_j^{NF}\odot\Phi_{k})\textbf{g}_{NF,j} \Big ).
\end{aligned}
\end{equation}
For $\mathcal{F}_{j}\in\left \{ \mathcal{F}_{j}^{NF},\mathcal{F}_{j}^{FF} \right \} $, the gradient at the $n$-th antenna position is given by
\begin{equation}\label{F_j_grad}
    \bigtriangledown_{\textbf{x}}\mathcal{F}_{j}= \frac{1}{2\sqrt{\textbf{b}_{j}^{H} \boldsymbol{\mathcal{H}}_{j}(\textbf{x})\textbf{b}_{j}}}\times \frac{\partial\textbf{b}_{j}^{H} \boldsymbol{\mathcal{H}}_{j}(\textbf{x})\textbf{b}_{j}}{\partial  \textbf{x}}= \frac{\bigtriangledown_{\textbf{x}}\mathcal{G}_{j,k}}{2\sqrt{\textbf{b}_{j}^{H} \boldsymbol{\mathcal{H}}_{j}(\textbf{x})\textbf{b}_{j}}} .
\end{equation}
For the Eve, following the same procedure, we can obtain
\begin{equation}
\begin{aligned}
&\bigtriangledown_{\textbf{x}}\mathcal{G}_{e,p}^{\mathrm{NF}}=\\
    &-\textbf{G}_{e}^{NF}\Big ( 2\left(\Upsilon_e^{NF}\odot\Theta_p\right) \textbf{g}_{NF,e}-2\left(\Upsilon_e^{NF}\odot\Phi_p\right)\textbf{w}_{NF,e} \Big )\\
&+\textbf{W}_{e}^{NF}\Big ( 2\left(\Upsilon_e^{NF}\odot\Theta_p\right) \textbf{w}_{NF,e}+2\left(\Upsilon_e^{NF}\odot\Phi_p\right)\textbf{g}_{NF,e} \Big ) ,
\end{aligned}
\end{equation}
\begin{equation}
\begin{aligned}
&\bigtriangledown_{\textbf{x}}\mathcal{G}_{e,p}^{\mathrm{FF}}=\\
    &-\textbf{G}_{e}^{FF}\Big ( 2\left(\Upsilon_e^{FF}\odot\Theta_p\right) \textbf{g}_{FF,e}+2\left(\Upsilon_e^{FF}\odot\Phi_p\right)\textbf{w}_{FF,e} \Big )\\
&+\textbf{W}_{e}^{FF}\Big ( 2\left(\Upsilon_e^{FF}\odot\Theta_p\right) \textbf{w}_{FF,e}-2\left(\Upsilon_e^{FF}\odot\Phi_p\right)\textbf{g}_{FF,e} \Big ) ,
\end{aligned}
\end{equation}
where $\Upsilon_e^{NF} =(\boldsymbol{\beta}_e^{\mathrm{NF}} (\boldsymbol{\beta}_e^{\mathrm{NF}})^H)$ and $\Upsilon_e^{FF} =(\boldsymbol{\beta}_e^{\mathrm{FF}} (\boldsymbol{\beta}_e^{\mathrm{FF}})^H)$. For $\mathcal{F}_{ej}\in\{\mathcal{F}_{ej}^{NF},\mathcal{F}_{ej}^{FF}\}$, we have
\begin{equation}\label{F_ej_grad}
    \bigtriangledown_{\textbf{x}}\mathcal{F}_{ej}= \frac{\bigtriangledown_{\textbf{x}}\mathcal{G}_{e,j}}{2\sqrt{\textbf{b}_{j}^{H} \boldsymbol{\mathcal{H}}_{e}(\textbf{x})\textbf{b}_{j}}} .
\end{equation}

\textit{Extension to the unified response (\ref{eq:unified response}):} Let $\psi_{n,j}=\omega_j\frac{2\pi}{\lambda}\left(r_j^{(n)}-r_j\right)-(1-\omega_j)\frac{2\pi}{\lambda}x_n\cos(\theta_j)$ be the phase of (\ref{eq:unified response}), so that $\mathbf{a}=\mathbf{g}_{\omega,j}-J\mathbf{w}_{\omega,j}$ with $g_{n,j}^{\omega}=\cos\psi_{n,j}$ and $w_{n,j}^{\omega}=\sin\psi_{n,j}$ (this reduces to the NF pair for $\omega_j=1$ and the FF pair with $w_{n,j}^{\omega}=-w_{n,j}^{FF}$ for $\omega_j=0$). Then $\mathcal{G}_{j,k}$ has the NF form (\ref{f(g,w) NF}) with $(\mathbf{g}_{\omega,j},\mathbf{w}_{\omega,j},\Upsilon_j(\omega_j))$, where $\Upsilon_j(\omega_j)=\boldsymbol{\beta}_j(\omega_j)\boldsymbol{\beta}_j(\omega_j)^H$. Its position gradient is obtained from the NF expression by replacing $\mathbf{G}_{NF,j}$ and $\mathbf{W}_{NF,j}$ with
\begin{equation}
    \mathbf{G}_{\omega,j}=\mathrm{diag}\left(\left\{\psi'_{n,j}\sin\psi_{n,j}\right\}_{n=1}^{N_t}\right),
\end{equation}
\begin{equation}
    \mathbf{W}_{\omega,j}=\mathrm{diag}\left(\left\{\psi'_{n,j}\cos\psi_{n,j}\right\}_{n=1}^{N_t}\right),
\end{equation}
where $\psi'_{n,j}=\frac{\partial\psi_{n,j}}{\partial x_n}=\frac{2\pi}{\lambda}\left(\omega_j\frac{x_n-r_j\cos(\theta_j)}{r_j^{(n)}}-(1-\omega_j)\cos(\theta_j)\right)$. The weight derivative needed in (\ref{eq:aperture gradient}) follows from $\partial\psi_{n,j}/\partial\omega_j=\delta_{n,j}$ in (\ref{eq:residual}):
\begin{equation}
\begin{aligned}
    \frac{\partial\mathcal{G}_{j,k}}{\partial\omega_j}=\mathbf{1}^T\Big[&-\mathbf{G}_{\delta,j}\Big(2(\Upsilon_j\odot\Theta_k)\mathbf{g}_{\omega,j}-2(\Upsilon_j\odot\Phi_k)\mathbf{w}_{\omega,j}\Big)\\
    &+\mathbf{W}_{\delta,j}\Big(2(\Upsilon_j\odot\Theta_k)\mathbf{w}_{\omega,j}+2(\Upsilon_j\odot\Phi_k)\mathbf{g}_{\omega,j}\Big)\Big],
\end{aligned}
\end{equation}
with $\mathbf{G}_{\delta,j}=\mathrm{diag}\left(\{\delta_{n,j}\sin\psi_{n,j}\}_{n=1}^{N_t}\right)$ and $\mathbf{W}_{\delta,j}=\mathrm{diag}\left(\{\delta_{n,j}\cos\psi_{n,j}\}_{n=1}^{N_t}\right)$. Since $\beta_n$ depends on both $x_n$ (in NF) and $\omega_j$, the terms $\frac{\partial\mathcal{G}_{j,k}}{\partial\beta_n}\frac{\partial\beta_n}{\partial x_n}$ and $\sum_n\frac{\partial\mathcal{G}_{j,k}}{\partial\beta_n}\frac{\partial\beta_n}{\partial\omega_j}$ must be included, where $\frac{\partial\mathcal{G}_{j,k}}{\partial\beta_n}=2g_{n,j}^{\omega}[\Theta_k\mathbf{p}-\Phi_k\mathbf{q}]_n+2w_{n,j}^{\omega}[\Theta_k\mathbf{q}+\Phi_k\mathbf{p}]_n$, $\mathbf{p}=\boldsymbol{\beta}_j\odot\mathbf{g}_{\omega,j}$, $\mathbf{q}=\boldsymbol{\beta}_j\odot\mathbf{w}_{\omega,j}$. These terms are small (relative order of $1/(kr_j)\approx10^{-4}$), but they ensure that the gradient is accurate. The same substitutions apply to $\mathcal{G}_{e,p}$, $\mathcal{F}_j$, and $\mathcal{F}_{ej}$, and the assembly of $\partial\Lambda/\partial\omega_j$ is the same as (\ref{eq:gradient1}).

\bstctlcite{IEEEexample:BSTcontrol}

\makeatletter
\apptocmd{\thebibliography}{\linespread{0.95}\selectfont}{}{}
\makeatother
\bibliographystyle{IEEEtran}
\bibliography{ref}

\end{document}